\documentclass[prd,aps,twocolumn,a4paper,showkeys,nofootinbib]{revtex4-1}
\usepackage{amsmath}
\usepackage{amsfonts}
\usepackage{amssymb}	
\usepackage{graphicx}
\usepackage{amsbsy}
\usepackage{color}
\usepackage{commath}
\usepackage{hyperref}
\allowdisplaybreaks

\usepackage{ulem} 

\def\M{{\cal F}}
\def\MM{\overline{{\cal F}}}
\def\mismatch{\MM}

\def\lm{{\ell m}}

\def\psilm{{\psi_4^\lm}}

\def\GRA{{\texttt{GR-Athena++}}}
\def\ETK{{\texttt{EinsteinToolkit}}}
\def\BAM{{\texttt{BAM}}}

\newcommand\etksim[1]{{\texttt{ETK{#1}}}}
\newcommand\athsim[1]{{\texttt{GRA{#1}}}}

\newcommand{\TEOB}[1]{\texttt{TEOBResumS{#1}}}

\newcommand\be{\begin{equation}}
\newcommand\ee{\end{equation}}

\usepackage{color}
\definecolor{cyan}{rgb}{0,0.9,0.9}
\definecolor{orange}{rgb}{0.9,0.5,0}
\definecolor{magenta}{rgb}{1,0,1}
\definecolor{purple}{rgb}{0.8,0.4,0.8}
\definecolor{gray}{rgb}{0.8242,0.8242,0.8242}
\definecolor{dodgerblue}{rgb}{0.12, 0.56, 1.0}
\definecolor{darkgrey}{rgb}{0.5,0.5,0.5}
\definecolor{darkgreen}{rgb}{0,0.65,0}

\begin{document}

\title{Towards numerical-relativity informed effective-one-body 
waveforms\\ for dynamical capture black hole binaries}

\author{Tomas \surname{Andrade}${}^{1}$}
\author{Juan \surname{Trenado}${}^{1}$}
\author{Simone \surname{Albanesi}${}^{2,3}$}
\author{Rossella \surname{Gamba}${}^{4}$}
\author{Sebastiano \surname{Bernuzzi}${}^{4}$}
\author{Alessandro \surname{Nagar}${}^{3,5}$}
\author{Juan \surname{Calderon Bustillo}${}^{6,7}$}
\author{Nicolas \surname{Sanchis-Gual}${}^{8}$}
\author{Jos\'e A.~\surname{Font}${}^{8,9}$}
\author{William \surname{Cook}${}^{4}$}
\author{Boris \surname{Daszuta}${}^{4}$}
\author{Francesco \surname{Zappa}${}^{4}$}
\author{David \surname{Radice}${}^{10,11,12}$}

\affiliation{${}^1$ Departament de F{\'\i}sica Qu\`antica i Astrof\'{\i}sica, Institut de
Ci\`encies del Cosmos, Universitat de
Barcelona, Mart\'{\i} i Franqu\`es 1, E-08028 Barcelona, Spain}
\affiliation{${}^2$ Dipartimento di Fisica, Universit\`a di Torino, via P. Giuria 1, 
10125 Torino, Italy}
\affiliation{${}^3$INFN Sezione di Torino, Via P. Giuria 1, 10125 Torino, Italy} 
\affiliation{${}^4$Theoretisch-Physikalisches Institut, Friedrich-Schiller-Universit{\"a}t 
Jena, 07743, Jena, Germany}  
\affiliation{${}^5$Institut des Hautes Etudes Scientifiques, 91440 Bures-sur-Yvette, France}
\affiliation{${}^6$Instituto Galego de F\'isica de Altas Enerx\'ias, Universidade de
Santiago de Compostela, 15782 Santiago de Compostela, Galicia, Spain}
\affiliation{${}^7$Department of Physics, The Chinese University of Hong Kong, Shatin, N.T., Hong Kong}  
\affiliation{${}^8$Departamento de Astronom\'ia y Astrof\'isica, Universitat de Val\`encia,
Dr.~Moliner 50, 46100 Burjassot (Val\`encia), Spain}
\affiliation{${}^{9}$Observatori Astron\`omic, Universitat de Val\`encia,
Catedr\'atico Jos\'e Beltr\'an 2, 46980 Paterna (Val\`encia), Spain}
\affiliation{$^{10}$Institute for Gravitation \& the Cosmos, The Pennsylvania State University, University Park PA 16802, USA}
\affiliation{$^{11}$Department of Physics, The Pennsylvania State University, University Park PA 16802, USA}
\affiliation{$^{12}$Department of Astronomy \& Astrophysics,  The Pennsylvania State University, University Park PA 16802, USA}

\begin{abstract}

Dynamical captures of black holes may take place in dense stellar media due to the emission of 
gravitational radiation during a close passage. 
Detection of such events requires detailed modelling, since their phenomenology
qualitatively differs from that of quasi-circular binaries. 
Very few models can deliver such waveforms, and none includes information from Numerical Relativity (NR)
simulations of non quasi-circular coalescences.
In this study we present a first step towards a fully NR-informed 
Effective One Body (EOB) model of dynamical captures. 
We perform 14 new simulations of single and double encounter mergers, and use this 
data to inform the merger-ringdown model of the \TEOB{-Dalì} approximant. We keep the initial energy  
approximately fixed to the binary mass, and vary the mass-rescaled, dimensionless angular momentum in the range $(0.6, 1.1)$, 
the mass ratio in $(1, 2.15)$ and aligned dimensionless spins in $(-0.5, 0.5)$. 
We find that the model is able to match NR to $97\%$, improving previous performances, 
without the need of modifying the base-line template. Upon NR informing the model, 
this improves to $99\%$ with the exception of one outlier corresponding to 
a direct plunge. The maximum EOB/NR phase difference at merger for the uninformed model is 
of 0.15 radians, which is reduced to 0.1 radians after the NR information is introduced. 
We outline the steps towards a fully informed EOB model of dynamical captures, and discuss 
future improvements.

\end{abstract}
\date{\today}

\maketitle

\section{Introduction}
\label{sec:introduction}

The detection and characterization of gravitational waves (GW) requires a multidisciplinary 
effort that combines instruments of exquisite precision -- the LIGO-Virgo-KAGRA (LVK) network 
of interferometers \cite{LIGOScientific:2014pky, VIRGO:2014yos, KAGRA:2020tym} -- and 
sophisticated data analysis techniques.
Thus far, the most numerous events in the LVK catalogue have been categorized as 
binary black holes (BBHs) with small or negligible eccentricity \cite{LIGOScientific:2021djp}. 
This type of event is expected to ensue from stellar binaries that have evolved in isolation 
from their environments and have radiated away their excess angular momentum, thus displaying quasi-circular orbits once they enter the sensitive band. 
However, some events do not easily fit in this category. In particular, the most massive BBH observed 
to date, GW190521~\cite{LIGOScientific:2020iuh, LIGOScientific:2020ufj}, has been shown to be
consistent with a dynamical capture of two nonspinning black holes~\cite{Gamba:2021gap}.
Alternative interpretations of this source are discussed in~\cite{Romero-Shaw:2020thy, 
Bustillo:2020syj, Gayathri:2020coq, Shibata:2021sau}. 

Dynamical captures may take place in dense stellar media such as globular clusters if the 
individual black holes radiate gravitational energy during a close passage~\cite{Rasskazov:2019gjw, Tagawa:2019osr}.
Since the phenomenology of these events is sensibly different from the
quasi-circular ones~\cite{Zevin:2018kzq,Samsing:2018isx,Nagar:2020xsk},
detailed modeling of the waveforms is required to detect and properly characterize such events through matched filtering. Indeed,
if quasi-circular templates are used for the search and analysis of waveforms generated by dynamical captures,
the events might be missed or incorrectly analyzed~\cite{East:2012xq,Loutrel:2020kmm,Nagar:2020xsk}.
Moreover, events of this kind can be detected for farther and heavier black hole systems~\cite{10.1111/j.1365-2966.2009.14653.x}. 
Dynamical captures are also relevant for next-generation detectors such as LISA~\cite{Amaro-Seoane:2018gbb} and
the Einstein Telescope~\cite{Punturo:2010zza}. The detectability rate for next-generation, ground-based detectors
has been estimated in Ref.~\cite{Mukherjee:2020hnm}.

Numerical Relativity (NR), i.e.~the fully fledged evolution of
Einstein's equations, provides the most detailed description of BBH
mergers. However, a systematic NR study of dynamical captures is presently missing. 
Low energy non-spinning encounters have been systematically studied by \citet{Gold:2012tk}. Fewer initial data including spin and mass ratio effects have been considered in Refs.~\cite{East:2012xq,Nelson:2019czq, Bae:2020hla,Gamba:2021gap}. 
NR simulations for hyperbolic encounters have also been recently considered in \cite{Jaraba:2021ces} including 
spin and varying mass ratios. 
Several new simulations of bound orbits but with large eccentricity
and precessing spins have also been recently reported~\cite{Gayathri:2020coq, Healy:2022wdn}.
The computational cost involved in spanning the possible orbital configurations for different binaries (mass ratio and spins) makes impractical to 
directly employ NR for a complete survey and for constructing waveform approximants.
Similarly to the circularized orbits case, our strategy is to exploit synergy with analytical relativity.
    
Reference \citet{Nagar:2020xsk} proposed the first analytical and complete general-relativistic description of the dynamics and waveforms of 
dynamical captures. 
The approach of \citet{Nagar:2020xsk} is based on the Effective One
Body (EOB) framework \texttt{TEOBResumS}~\cite{Damour:2014yha, Bernuzzi:2014owa, Nagar:2017jdw, Nagar:2018zoe, 
Akcay:2018yyh,Nagar:2019wds,Nagar:2020pcj} and in particular on the generic orbits version 
\TEOB{-Dalì}~\cite{Chiaramello:2020ehz, Nagar:2021gss, Nagar:2021xnh}.
This model was the basis for the analysis of GW190521 reported
by~\cite{Gamba:2021gap}, as well as the eccentric analyses presented in~\cite{Iglesias:2022xfc}. It produces a quantitative predictions for the waveform
from the entire orbital parameter space, including spin and mass-ratio effects.
Remarkably, despite not being informed by any eccentric NR simulation, \TEOB{-Dalì} 
shows mismatches below $1\%$ for almost all the available mildly eccentric configurations 
of the SXS catalog~\cite{Nagar:2021gss} as well as for the RIT catalog~\cite{Gamba:2021gap}.
The accuracy of the EOB dynamics has been also tested in the hyperbolic encounter scenario by comparing the 
EOB/NR scattering angles for both non-spinning and spinning configurations~\cite{Damour:2014afa,Hopper:2022rwo,Damour:2022ybd}.
Moreover, the accuracy of the fluxes, i.e. the non-conservative part of the dynamics, 
has been tested both for comparable mass and high-mass ratio systems~\cite{Albertini:2021tbt,Albertini:2022rfe,Albertini:2022dmc}
in the quasi-circular case, and for hyperbolic and eccentric systems with extreme mass ratio~\cite{Albanesi:2021rby,Albanesi:2022ywx}.
However, even if a few EOB/NR comparisons for equal mass binaries have been reported in the Supplemental
Material of Ref.~\cite{Gamba:2021gap}, the dynamical capture scenario for comparable mass black holes still needs to be
explored in detail. 
A search for hyperbolic encounters using public data was carried out in \cite{Morras:2021atg}, with results consistent with the false alarm search rate. Observational implications of dynamical captures have recently been considered in \cite{Ebersold:2022zvz}.

The main goal of this work is to assess the capabilities of the latest version of \TEOB{-Dalì} \cite{Nagar:2021gss} in the 
dynamical capture scenario.
We carry out a series of new NR simulations of dynamical captures, quantitatively verify some of the key predictions of \citet{Nagar:2020xsk}, and propose a first strategy to inform the model with NR data. 
As we show below, we find that by appropriately informing our model with data coming from NR simulations
we are able to obtain waveforms more than $99\%$ faithful to NR in most cases. 

This paper is organized as follows:  
In Sec.~\ref{sec:NR} we present our new NR simulations and perform the first comparison 
between the {\ETK} and \GRA~ NR codes. We describe in some detail the configurations chosen, 
the two codes employed to obtain them and the tests that we performed to ensure convergence. 
We also discuss the post-processing of the NR data, with a focus on waveform extraction techniques.
In Sec.~\ref{sec:EOB} we briefly introduce the \TEOB{-Dalì} model, discuss the 
NR-informed parameters that we consider and tackle the issue of connecting NR and EOB initial data.
In Sec.~\ref{sec:eobnr} we perform comparisons between our simulations and \TEOB{-Dalì}, presenting both time-domain phasing comparisons and mismatch computations.
Finally, in Sec.~\ref{sec:conclusions} we summarize and discuss our results.
Geometric units with $G = c = 1$ are employed throughout 
{the text}, unless otherwise specified.

\section{Numerical relativity simulations}
\label{sec:NR}

\subsection{Numerical codes}
\label{sec:num_codes}
\subsubsection{Einstein Toolkit}

For the bulk of this work we use the open source software {\ETK}~\cite{Loffler:2011ay}. 
This code is organized as a main driver (Cactus) and a series of modules (thorns) 
that perform specific tasks.
BBH initial data are computed with the \texttt{TwoPunctures} thorn~\cite{Brandt:1997tf, Ansorg:2004ds}, 
which is appropriate to solve the constraint equations on the initial $t=0$ slice for 
the modest values of spins and mass ratio considered in this work. 
In the range of parameters we have explored, this algorithm converges to the desired 
numerical precision within a few 
iterations. 
The time evolution is carried out by the \texttt{MLBSSN} thorn which implements the 
BSSN formulation of vacuum General Relativity \cite{Baumgarte:1998te, Shibata:1995we}. 
We use 6th order finite differences for spatial derivatives and a method-of-lines 
time integration with a 4th order Runge-Kutta scheme.
Moreover, we include Kreiss-Oliger dissipation of order 9 scaled with a factor of $\epsilon = 0.1$. The CFL factor for our \ETK{} runs is $0.1$. 

Our grid setup consists of a cubic box of edge $L = 640 M$ (in the coarsest level, 
the cartesian coordinates range from $-320 M$ to $320 M$)
and spacing at the coarsest level $\delta x = \{3, 4, 6 \} M$, corresponding to (Low, Medium, High) resolutions. 
Here $M$ is the total mass of the binary, defined as the sum of the individual ADM masses.
We have carried out most of our simulations at $\delta x = 4 M$. We also provide 
convergence tests for selected simulations consider the low and high resolutions. 
The code uses Berger-Oliger mesh refinement with 9 overlapping refinement 
levels in a box-in-box fashion, with the two most refined boxes containing the two punctures. 
The boxes follow the two punctures, whose position is determined 
during the evolution with the \texttt{PunctureTracker} thorn. 
This setting results in a puncture resolution of 
$\delta x_p = 4/2^8 M = 0.015625 M$.
For unequal masses, we add an extra refinement level {only} 
at the smaller black hole. This was proven crucial to obtain robust results 
as we vary resolution. 
We exploit the reflection symmetry of our configurations to reduce the computational 
domain by a half in the $z-$direction. 

We extract the wave content of our simulations using the thorns \texttt{WeylScal4} 
and \texttt{Multipole} which outputs the Weyl scalars $\psi_4$ expanded 
in spherical harmonics up to $\ell = 8$, at fixed radius.
We check that $ R \psi_4 = {\rm const}$, as expected for 
the extraction radii $R = \left\lbrace 70,\,80,\,90,\,100,\,110\right\rbrace \,M$, which are 
located in a refinement level at the same resolution 
$\delta x = 4/2^{2} M = 1 M$. On the contrary, extraction radii 
$R \geq 120$ fall within levels of lower resolution, which 
causes power loss due to numerical dissipation. 

\subsubsection{GR-Athena++}
\label{sec:gra}
In this work we also extend the set of {\GRA}~\cite{Daszuta:2021ecf} simulations presented 
in~\cite{Gamba:2021gap}.
The initial data are computed with a stand-alone version of the thorn 
\texttt{TwoPunctures}~\cite{Brandt:1997tf, Ansorg:2004ds}. 
The time evolution is then performed by {\GRA} using the 
Z4c formulation~\cite{Bona:2003fj}. 
Moving puncture gauge conditions with the same parameters in Ref.~\cite{Daszuta:2021ecf} are  adopted.
As in the {\ETK} case, we use 6th order finite-difference 
methods for the spatial derivatives and we perform the time-integration using a 4th order 
Runge-Kutta algorithm. In contrast, for the simulations performed here, Kreiss-Oliger dissipation of order $8$ is included with a factor $\epsilon =0.02$.
We also choose a bigger 
box with edge $L=3072 M$, so that in the coarsest level the cartesian coordinates range from $-1536 M$ to $1536 M$.
The AMR in 
\GRA{} is oct-tree based, with the grid organized as an initial
Mesh divided into Meshblocks which have all the same number of
grid points but (possibly) different physical size. For a cubic
initial Mesh and cubic Meshblocks, the
grid setup in {\GRA} is regulated by three parameters: the number of 
grid points in the edges of the unrefined initial mesh
$N_M$, the number of grid points 
in the edges of Meshblocks $N_B$, and the number of physical 
refinement levels $N_L$. The grid structure is ultimately
determined by an AMR criterion, which, when satisfied, (de)refines a given MeshBlock, resulting in a (larger) smaller block with (half) double the resolution. For BBH simulations
the AMR criterion used mimics the box-in-box strategy mentioned above.
In our simulations we consider $N_B=16$, $N_L=11$, and $N_M = \left\lbrace 128,\,192,\,256\right\rbrace$.
These values are chosen so that the resolution at the punctures 
is the same as the simulations performed with the \ETK{} code. For all the runs we consider 
a CFL number of 0.5
The Weyl scalar is then extracted at $R=\left\lbrace 80,90,100,110,120,130,140 \right\rbrace M$ 
using an approximate geodesic sphere built 
using 9002 vertices. For the three grids considered, the resolutions in the extraction zones at the merger time 
are $\delta x_R = \left\lbrace 3,2,1.5 \right\rbrace M$ for $R>96 M$, while they are $\delta x_R = \left\lbrace 1.5,1,0.75 \right\rbrace M$ 
for $48 M< R < 96M$. Note that at the beginning of the simulation the extraction zone for $R>96M$ 
does not have a uniform resolution since the positions of the two punctures make some portions 
of the zone more refined.
We observed that $R \psi_4$ remains approximately constant at all the extraction radii, 
showing that $\psi_4$ scales as $1/R$, as expected. 
%
More technical details on the structure of the grid and of the geodesic sphere can 
be found in Ref.~\cite{Daszuta:2021ecf}. 
Note that no grid-symmetries are employed for these runs.

\subsection{Initial data}

\begin{figure}[t] 
\centering
\includegraphics[width=.5\textwidth]{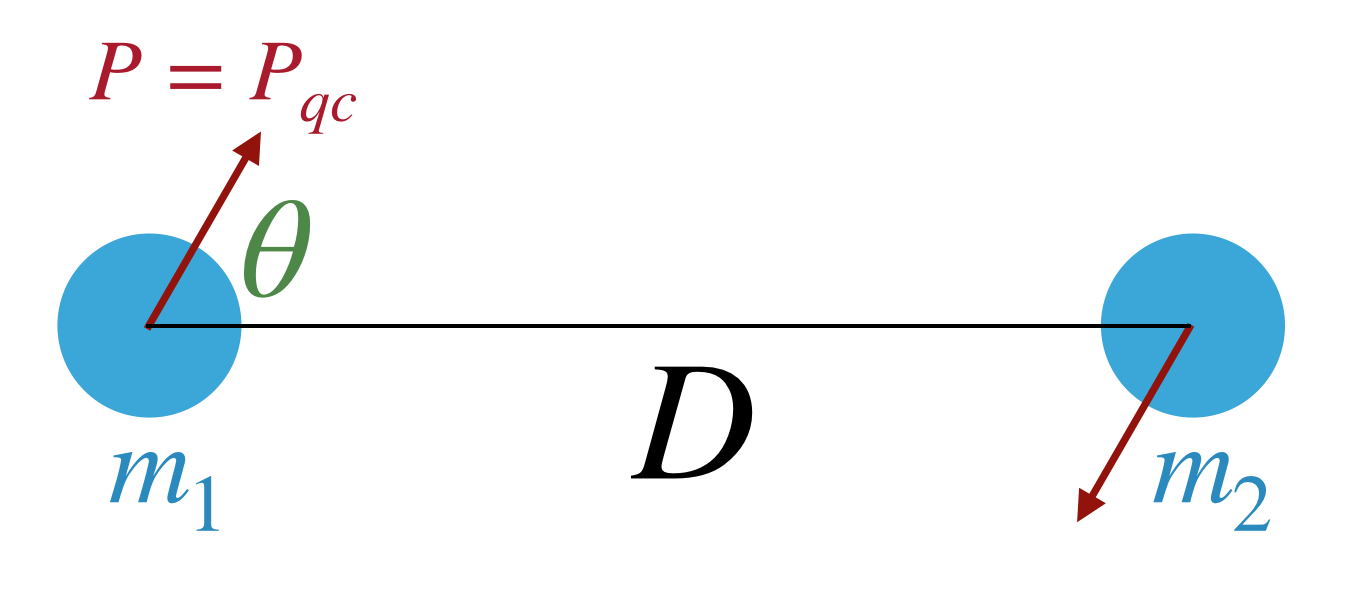}
\caption{Schematic depiction of our initial data. The initial
  separation $D$ and the the value of $P$ are kept fixed at $20 M$ and
  $0.06175M$, respectively, and we vary the initial angle $\theta$ as
  well as the intrinsic spins of the BH components.} 
\label{fig:ic}
\end{figure}

We consider initial data consisting of two black holes of
(quasi-local) ADM masses $M_1$ and $M_2$ separated by
a coordinate distance $D$. The total mass of the binary is $M = M_1 + M_2$. 
We take their ADM linear momenta to be $\vec P_1 = -
\vec P_2$, with $\vec P_1 = P_{qc}(\cos \theta, \sin \theta, 0)$ where
$P_{qc}$ corresponds to the quasi-circular value.  
The black holes have spins $\vec{J}_1$, $\vec{J}_2$ aligned with the
orbital angular momenta, so that $\vec J_1 = \chi_1 M_1^2 (0,0,1)$,
$\vec J_2 = \chi_2 M_2^2 (0,0,1)$. Here $\chi_{1,2}$ are dimensionless 
spin parameters. 
Following \cite{Gold:2012tk} we take $D = 20 M$ which in turn implies $P_{qc} = 0.06175 M$.  
See Fig \ref{fig:ic} for a schematic depiction of our initial data.

\subsection{Simulation results}
\label{sec:sim_results}
\begin{table*}[t]
\caption{\label{tab:ETK_sims}Dynamical capture BBHs configurations considered in this work.
From left to right the columns report: the configuration name; the mass ratio $q=m_1/m_2$, the 
angle  $\theta$ of the component of the tangential momentum; the dimensionless spins $\chi_i\equiv S_i/m_i^2$,
the initial ADM energy and angular momentum, $(E_0^{\rm ADM},J_0^{\rm ADM})$, the final energy and angular momentum $(E_f,J_f)$
and the gravitational wave losses $(\Delta E,\Delta J)$, all expressed in units of the total mass $M$. 
The initial separation is $D=20M$ for all configurations. All configurations are simulated using  the \ETK{}.
Those marked with a star also with the \GRA{} code. See Table \ref{tab:self_conv_info} below for additional information.
}
\begin{center}
\begin{ruledtabular}
\begin{tabular}{ c c | c c c c c c c c c  c} 
\# & ID & $q$  & $\theta$ [deg] &  $\chi_1$  &  $\chi_2$  &  $E_0^{\rm ADM}/M$  &  $J_0^{\rm ADM}/M^2$ & $E_f/M$ & $J_f/M^2$ &  $\Delta E_{EOB}/M$ &  $\Delta J_{EOB}/M^2$   \\ 
 \hline 
 \hline 
1 &\etksim{37q1s0} & 1.00 &  37 &  0.00 &  0.00 &  0.99427 &  0.74321 &  0.97812 &  0.65522 &  0.00000 &  0.00446 \\ 
2 &\etksim{42q1s0} $^*$ & 1.00 &  42 &  0.00 &  0.00 &  0.99427 &  0.82634 &  0.96521 &  0.67555 &  0.00005 &  0.00488 \\ 
3 & \etksim{44q1s0} & 1.00 &  44 &  0.00 &  0.00 &  0.99428 &  0.85786 &  0.95755 &  0.66749 &  0.00000 &  0.00515  \\ 
4 &\etksim{46q1s0} & 1.00 &  46 &  0.00 &  0.00 &  0.99428 &  0.88834 &  0.94903 &  0.64455 &  0.00008 &  0.00486  \\ 
5 & \etksim{48q1s0} $^*$ & 1.00 &  48 &  0.00 &  0.00 &  0.99428 &  0.91774 &  0.95035 &  0.60959 &  0.00009 &  0.00539 \\ 
6 & \etksim{50q1s0} $^*$ & 1.00 &  50 &  0.00 &  0.00 &  0.99429 &  0.94602 &  0.94858 &  0.62682 &  0.00025 &  0.00094  \\ 
\hline
7 & \etksim{42q1s025{--}}& 1.00 &  42 &  $-0.25$ &  $-0.25$ &  0.99425 &  0.70134 &  0.97564 &  0.59586 &  0.00000 &  0.00421  \\ 
8 & \etksim{42q1s025{++}}& 1.00 &  42 &  0.25 &  0.25 &  0.99433 &  0.95134 &  0.94320 &  0.70570 &  0.00000 &  0.00571  \\ 
9 & \etksim{42q1s050{--}} & 1.00 &  42 &  $-0.50$ &  $-0.50$ &  0.99439 &  0.57634 &  0.98122 &  0.49612 &  0.00000 &  0.00346  \\ 
10 & \etksim{42q1s050{++}} & 1.00 &  42 &  0.50 &  0.50 &  0.99454 &  1.07634 &  0.93980 &  0.73469 &  0.00016 &  0.00606  \\ 
11 &\etksim{42q1s050{+-}} & 1.00 &  42 &  0.50 &  $-0.50$ &  0.99447 &  0.82634 &  0.96517 &  0.67477 &  0.00000 &  0.00496  \\ 
\hline
12 & \etksim{42q1.5s0} & 1.50 &  42 &  0.00 &  0.00 &  0.99501 &  0.82634 &  0.96229 &  0.65471 &  0.00000 &  0.00496  \\ 
13 & \etksim{42q2s0} & 2.00 &  42 &  0.00 &  0.00 &  0.99641 &  0.82634 &  0.95710 &  0.58005 &  0.00079 &  0.00444  \\ 
14 &\etksim{42q2.15s0} & 2.15 &  42 &  0.00 &  0.00 &  0.99687 &  0.82634 &  0.96340 &  0.58504 &  0.00012 &  0.00180 \\
\end{tabular}
\end{ruledtabular}
\end{center}
\end{table*}

We considered 14 configurations that were simulated using {\ETK}. 
Table~\ref{tab:ETK_sims} collects the initial physical parameters, while
in Appendix~\ref{app:trajectories} we show the puncture trajectories and
the time evolution of the $\ell=m=2$ multipole of the Weyl scalar $\psi_4$.
Simulations will be refereed to using their ID as defined in the first column of 
Table \ref{tab:ETK_sims}. The phenomenology of the transition from eccentric
inspiral, zoom-whirl behavior and dynamical capture was studied in 
Ref.~\cite{Pretorius:2007jn,Gold:2012tk} as the initial angle $\theta$ is varied. 
For angles close to $\theta=0$ (head-on collision) one has direct plunges.
For fixed values of the initial separation $D$, one can have various
close encounters (including zoom-whirl behavior) before the final merger.
Finally, for angles $\theta$ beyond a certain threshold, the black hole do not merge
and follow hyperbolic trajectories.
In our set of simulations, for equal masses and $\theta$ below 48 degrees we find plunges, 
in \etksim{37q1s0,42q1s0,44q1s0,46q1s0}, while \etksim{48q1s0} begins to display features 
of a zoom-whirl which can be fully appreciated in \etksim{50q1s0}.
Simulations \etksim{42q1s0,48q1s0,50q1s0} correspond to some of the configurations presented in 
Ref.~\cite{Gold:2012tk}. We find good qualitative and quantitative agreement between
puncture tracks and waveforms. These are a plunge, a transition to double encounter 
and a double encounter, respectively. 
Note that this agreement is particularly non-trivial for \etksim{48q1s0}, since, due to the fact
that this is a boundary case between single and double encounters, slight changes 
on the initial data or resolution can lead to different results. 
We further analyze these three cases by performing self-convergence tests 
and a code comparison between \ETK{} and \GRA{}, see Sec.~\ref{sec:NRconsistency}.

As also stated in \cite{Gold:2012tk}, varying the mass ratio can also 
result in zoom-whirls. We explore this regime in our series \etksim{42,42q150,42q200}, 
\etksim{42q2.15s0}, going from a plunge to a fully developed zoom-whirl. 
Moreover, we have observed that zoom-whirls can be induced by increasing the angular 
momenta by adding spin to the components of a binary which yields a plunge orbit in the 
non-spinning case. This is found in our simulations \etksim{42q1s050--, 42q1s025--, 
42q1s0, 42q1s025++, 42q1s050++} as discussed in Sec,~\ref{sec:spin} and is analytically
explained via the spin-orbit interaction.
NR simulations are also known to suffer from ``junk radiation'' i.e. radiation caused 
by the underlying conformally flat assumption used for the computation of the initial data.
In our case, we observe a small burst of radiation in $\psi_4$ which arrives at 
the extraction radius $R$ at a time $t_{\rm junk} \sim R$. 
We discuss in more detail the impact of junk radiation on our simulations 
in Sec. \ref{sec:energetics} below.

\subsection{Post-processing}
\label{sec:post-processing}

The Weyl scalar $\psilm$ that is given in output in the numerical simulations that we consider is extracted at a finite radius $R$ and thus needs to be extrapolated at infinity. In this work we consider the extrapolation proposed in Refs.~\cite{Lousto:2010qx,Nakano:2015rda},
\begin{equation}
\label{eq:extrapolation}
\lim_{r \rightarrow \infty} r \psilm \simeq A 
\bigg( r \psilm - \frac{(\ell-1)(\ell+2)}{2 r} \int dt \, r \psilm \bigg)
\end{equation}
where $A=1-2M/R$ and $r=R (1+M/(2R))^2$.
The plus and cross polarizations of the strain can be expanded in spin-weighted harmonics as
\begin{equation}
h_+(\Omega, t) - i h_\times (\Omega, t) = \frac{1}{R} \sum_\lm\, _{-2}Y_{\lm}(\Omega)  h_{\lm} (t)
\end{equation}
\noindent where $\Omega$ is the angular dependence. 
The corresponding waveform and fluxes can be obtained from the extrapolated scalar since
\begin{equation}
\label{eq:psi4_hlm}
R\psilm = \ddot{h}_\lm.
\end{equation}
It is well-known that performing this double time integration is subtle due to the presence of numerical noise which induces drifts in the signal \cite{Reisswig:2010di}. Note moreover that the algorithm in \eqref{eq:extrapolation} requires an additional integral, so extraction of the extrapolated strain modes takes three integrals in total.
In the case of circularized binaries, it is well established that the most reliable procedure to follow is the fixed-frequency integration (FFI), where the integration is performed in 
the frequency-domain and a frequency cut-off $\omega_0$ is introduced to 
get rid of the unphysical features \cite{Reisswig:2010di}. In that case, since the orbits are quasi-circular,
the frequency of the emitted gravitational waves is a monotonic increasing function of 
the time, and therefore it is straightforward to identify the value of $\omega_0$.
However, in the case of non-circularized binaries, and in particular for dynamical
capture, it is not clear how to identify the cut-off. 
In particular, we observe that for choices of cut-offs which are large enough to remove the drift in the ring-down,  FFI integration makes the amplitude of the precursor unphysically small. 
To overcome this issue, we use a time-domain integration and then remove the drift in the resulting signals.

For the leading $(2,2)$ modes, after each time integration (including the one in 
\eqref{eq:extrapolation}) we remove a complex constant by fitting a 0-th order polynomial 
in a $100 M$ interval after the maximum of $\psi_4$.  
For higher modes, the noisier signals require a more elaborate subtraction. 
In this case, the first integrals of $\psi_4$ when treated as above present a 
small drift, which is significantly amplified by performing the second integral. 
We eliminate the drift from the final signal by performing a 5th order polynomial fit extracted from the whole signal after junk radiation has finalized. 

We obtain the strain modes by extracting $\psi_4$ at $R = 100 M$, extrapolating to infinity with \eqref{eq:extrapolation} and computing the double time integral using direct time integration and subtracting the drift as explained above. We have observed that extrapolation in conjunction with direct time integration can be delicate for some signals and extraction radii. Our choice of extraction at $R = 100 M$ is the one that appears most robust.

When visualizing our signals in the time domain we will employ the retarded time $t - r_*$, with $r_*$ the tortoise coordinate
\begin{equation}
\label{eq:r_start}
	r_* = R + 2 M \log [R/(2M) - 1]
\end{equation}
\noindent with $R$ the extraction radius and $M$ the sum of the individual ADM masses.

\begin{figure*}[t]
  \includegraphics[width=.49\textwidth]{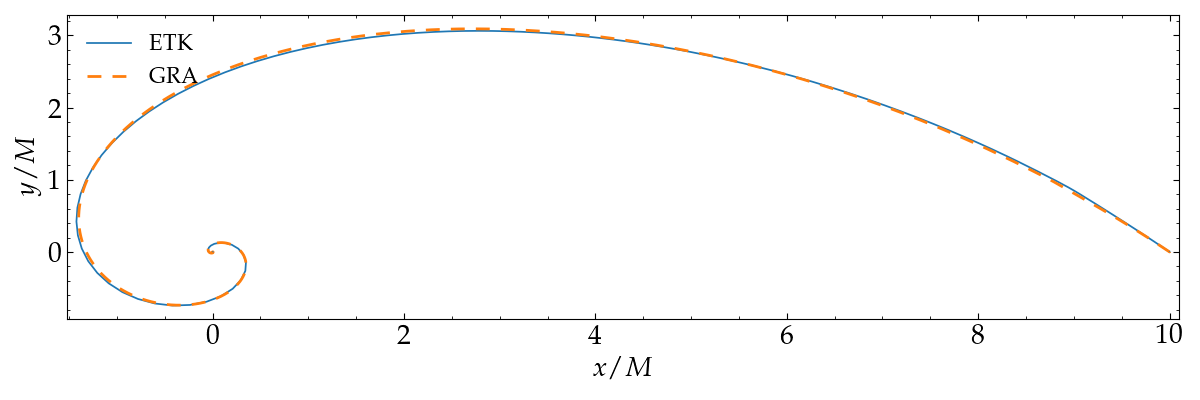}  
  \includegraphics[width=.49\textwidth]{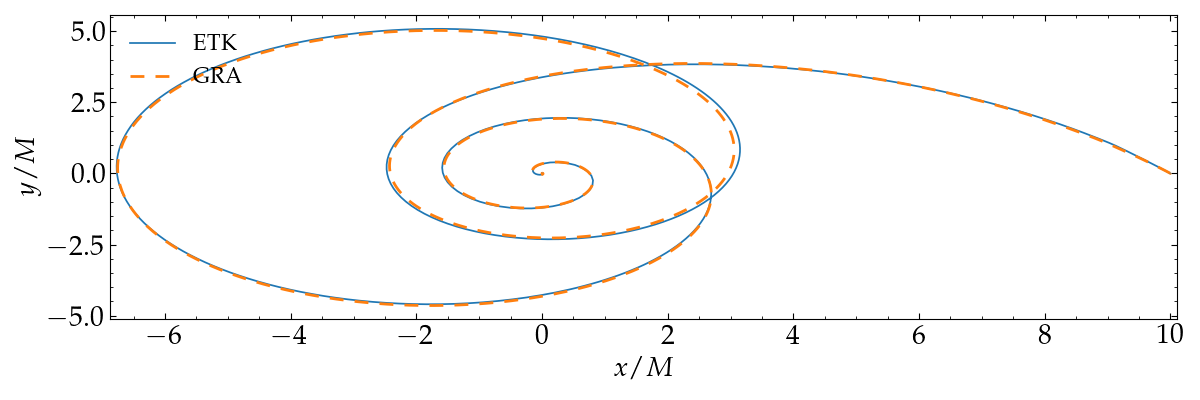}  
  \includegraphics[width=.49\textwidth]{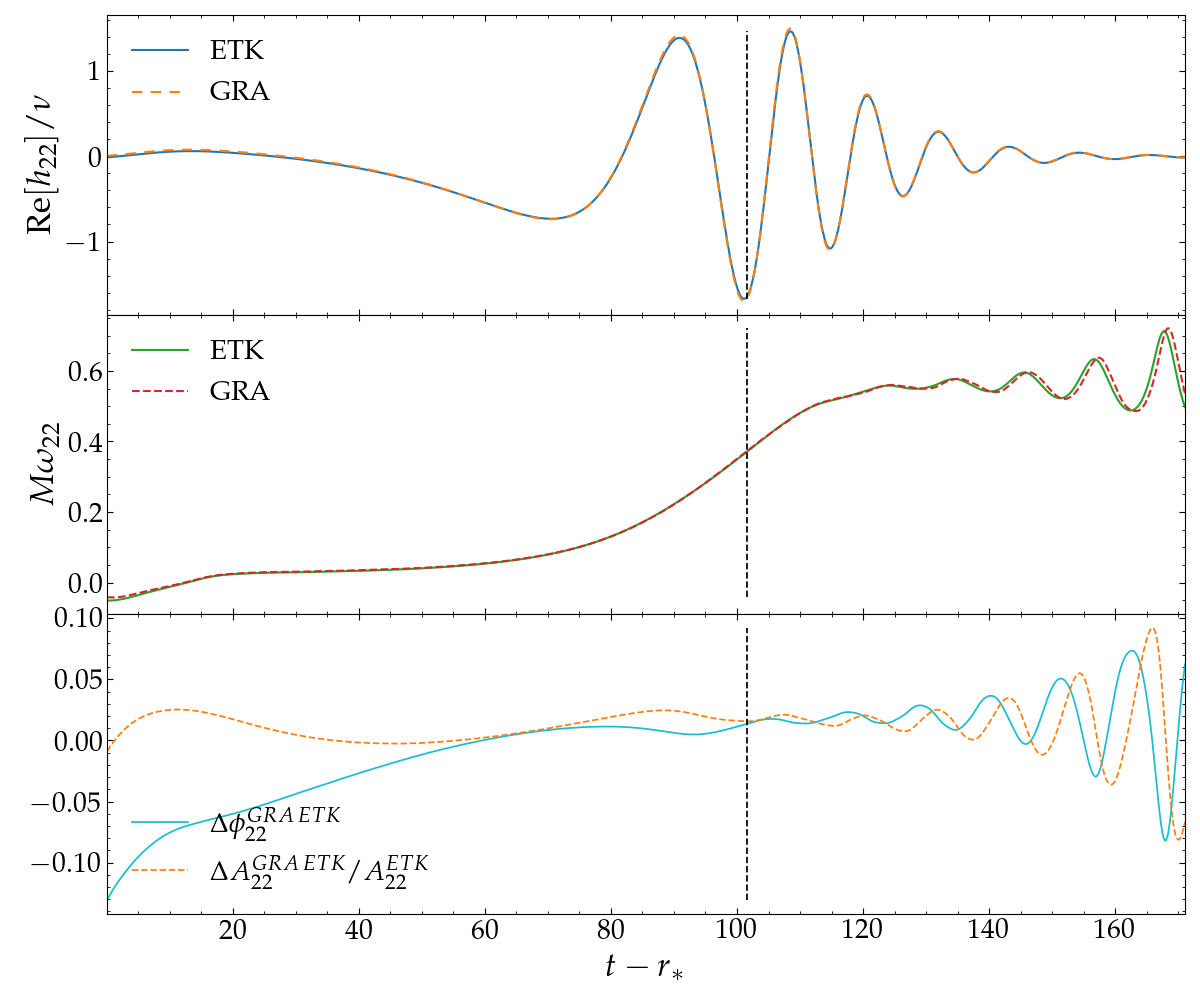}  
  \includegraphics[width=.49\textwidth]{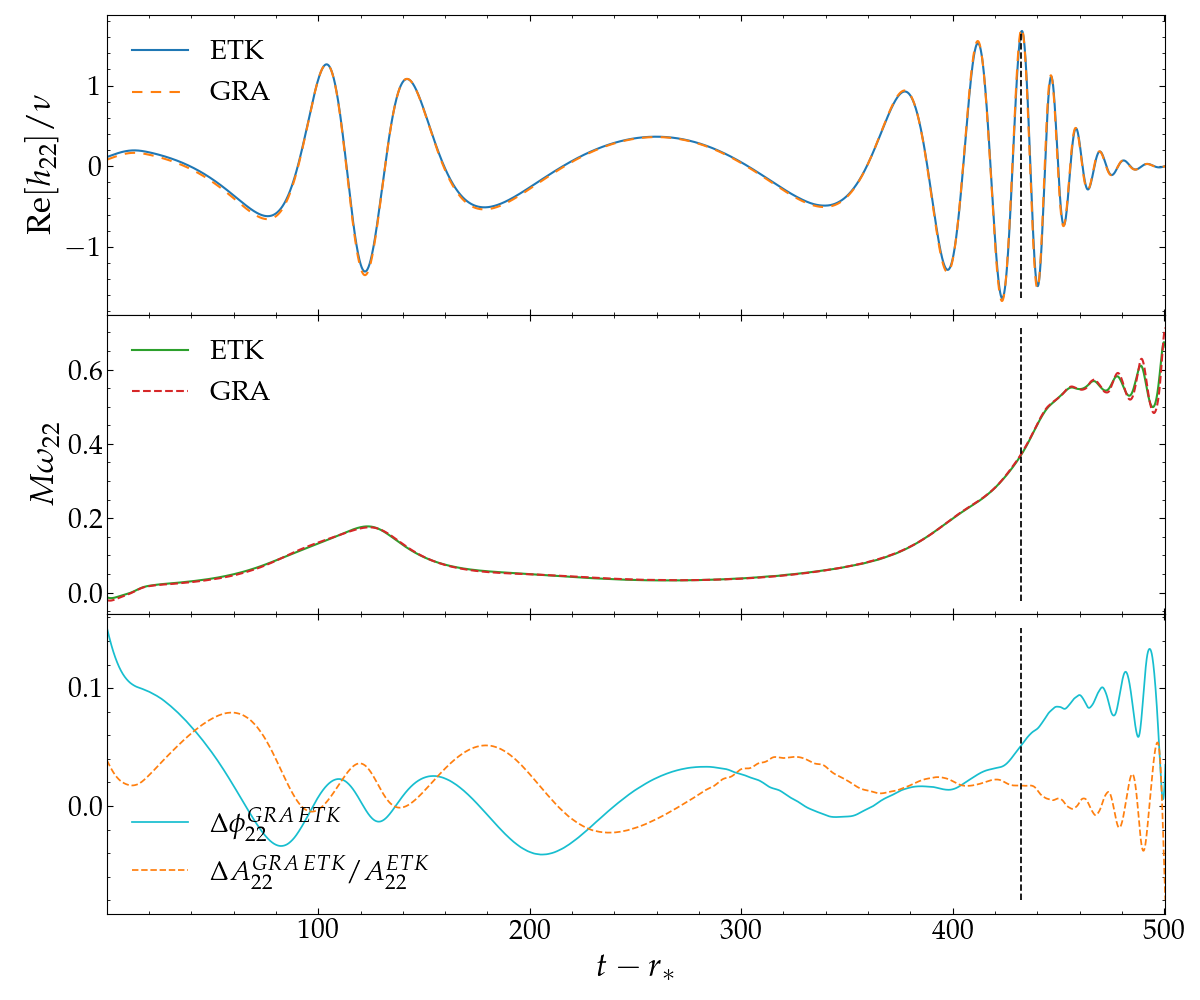}  
  \caption{\label{fig:nr_sample_42_50} Example orbit, waveforms, frequency, and amplitude and phase difference, for a direct capture 
    (\etksim{42q1s0}, \athsim{42q1s0}, left panel) and zoom-whirl (\etksim{50q1s0}, \athsim{50q1s0}, right panel). 
The differences are defined in \eqref{eq:delta_codes}.
We only show the trajectories of one of the black holes to ease visualization. Since these are equal-mass simulations the other trajectory can be obtained by reflection symmetry.
In the direct capture case the black holes plunge almost immediately, without completing a full orbital cycle. Therefore, the waveform is dominated by merger and ringdown. Conversely, in the zoom-whirl scenario, the bodies undergo multiple encounters before merging, each close passage corresponds to a GW burst.
In our code comparison, there is no alignment applied to the waveforms. We show the merger time as a vertical dashed on bottom panels. 
}
\end{figure*}

We display the results of our post-processed waveforms for \etksim{42q1s0,50q1s0} in 
Fig.~\ref{fig:nr_sample_42_50}. We write the modes of the strain in terms of the amplitude and phase as
\begin{equation}
\label{eq:strain_A_phi}
    h_{\ell m}(t) = A_{\ell m}(t) e^{i \phi_{\ell m}(t)}
\end{equation}
and introduce the difference between \ETK{} and \GRA{} results by
\begin{align}
\label{eq:delta_codes}
\nonumber
	\Delta A_{22}^{GRA \, ETK} &:= A_{22}^{GRA} - A_{22}^{ETK} , \\
	\Delta \phi_{22}^{GRA \, ETK} &:= \phi_{22}^{GRA} - \phi_{22}^{ETK}
\end{align}

\subsection{Energetics}
\label{sec:energetics}

We compute the radiated energy and angular momentum as a function of time, as (see e.g. \cite{Damour:2011fu})
\begin{align}
    \dot E =  \frac{1}{16 \pi} \sum_{(\ell, m)} \dot h_{\ell m}  \dot h_{\ell m}^* \\
    \dot J =  \frac{1}{16 \pi} \sum_{(\ell, m)} m \Im [  h_{\ell m} \dot h_{\ell m} ^*]
\end{align}
Note that here we are formally extrapolating to infinity by using the formula \eqref{eq:extrapolation}. 
Our data includes modes with $2 \leq l \leq 8$ and $m = -\ell \ldots \ell$, so in practice the sums are limited to these values. We have found that the modes 
$m = 0$ are numerically noisy so do not include them in the computation. 
Integrating the fluxes $\dot E$, $\dot J$ over time, we obtain the total radiated 
energy and angular momentum. This allows us to define the energy and angular momentum as a 
function of time $t$
\begin{align}
\label{eq:EJ}
    {\cal E}(t) &= E_i^{\rm ADM} - \int_{t}^\infty dt' \dot E(t') \\
    {\cal J}(t) &= J_i^{\rm ADM} - \int_{t}^\infty dt' \dot J(t')
\end{align}
\noindent where $E_i^{\rm ADM}$, $J_i^{\rm ADM}$ are the initial ADM total energy and angular momentum 
of the system as shown in Table \ref{tab:ETK_sims}.
Following \cite{Nagar:2015xqa}, we define the dimensionless binding energy and dimensionless angular momentum as
\begin{align}
\label{eq:Eb_and_j}
    E_b(t) &=  \frac{{\cal E}(t) - M}{\mu} \\
    j(t) &= \frac{{\cal J}(t)}{M \mu}
\end{align}
\noindent where $M = m_1 + m_2$ is the sum of the individual initial ADM masses and $\mu = m_1 m_2/M$.

Energetics are a meaningful and robust tool to compare NR data to EOB,
e.g.~\cite{Damour:2011fu,Bernuzzi:2012ci,Bernuzzi:2013rza,Nagar:2015xqa}. 
In the case of quasi-circular orbits, the NR initial puncture
parameters can be constructed to match the 3PN prediction.
When junk radiation is correctly taken into account, the initial
evolution drives the binary very close to the EOB curve, that is more
bound than the 3PN curve~\cite{Damour:2011fu}.
We observe in our dynamical capture
scenario that the initial burst of junk radiation has a smaller impact
on the energetics with respect to the quasi-circular case.
However, other sources of inaccuracies like residual gauge ambiguities
in the determination of the puncture parameters from the EOB
energetics and the reconstruction of the strain from $\psi_4$ can play
a role. Overall, we observe these effects are sufficiently small 
that the EOB/NR agreement is within the NR errors in the early phase of
the dynamics. 
However, in some cases we allow for a small (less than $1 \%$) adjustment in the 
ADM quantities to obtain EOB waveforms that
best match the NR simulations. These corrections are then also
employed when we compare the energy curves between EOB and NR.

\subsection{Faithfulness}
\label{sec:match_def}

The faithfulness (or match) is a common measure to quantify the global difference between 
two waveforms. For two signals in the time domain $h_1(t)$, $h_2(t)$ 
the match is defined as 
\begin{equation}
\label{eq:F}
    \M = \frac{\langle h_1, h_2 \rangle}{\sqrt{\langle h_2, h_2 \rangle \langle h_1, h_1 \rangle}}\,,
\end{equation}
where the inner product, assuming a noise function $S_n(f)$, is defined as 
\begin{equation}
    {\langle h_1, h_2 \rangle} = 4 \Re \int \frac{\tilde{h}_1(f) \tilde{h}^*_2(f)}{S_n(f)} df
\end{equation}
Here the $\tilde{h}_1(f)$ and $ \tilde{h}_2(f)$ are the Fourier transforms of 
the time domain waveforms, and the integral covers the frequency interval $Mf\in[0.005, 0.1]$ 
of the signal. 
In this work we consider a uniform PSD ($S_n(f)\equiv1$) unless explicitly stated.
We have checked that the power is concentrated in this interval for all of our 
simulations. In turn, the unfaithfulness, or mismatch, is given by 
$\MM = 1 - \M$.
The unfaithfulness, which ranges in [0,1], is equal to the fractional loss of 
signal-to-noise ratio due to the difference between the two compared waveforms.
We compute the matches using the algorithm \texttt{optimized$\_$match} 
implemented in {\tt pyCBC}, which efficiently aligns the waveforms optimizing 
over the differential phases and time shifts.

\subsection{Consistency of the numerical results} 
\label{sec:NRconsistency}

We discuss the consistency of the results obtained with \ETK and \GRA, focusing on simulations \etksim{42q1s0, 48q1s0, 50q1s0}. 
This will provide error estimates for our NR simulations needed to assess the EOB/NR comparisons later on.

We consider three resolutions, resulting from taking $\delta x = \{3, 4,6 \} M$ at the coarsest level for {\ETK} and 
$N_M = \{256, 192, 128\}$ for \GRA, since their corresponding puncture resolutions match. 
We summarize the relevant technical information for these simulations in Table \ref{tab:self_conv_info}. In the main text, we focus on the amplitude and phase differences for the leading 
modes $(2,2)$ of the strain, and comment on some other observables in the appendices.

\begin{table}[h]
\caption{\label{tab:self_conv_info} [Technical information for simulations \etksim{42q1s0,48q1s0, 50q1s0} and 
\athsim{42q1s0,48q1s0, 50q1s0}. 
We have runs at Low (L), Medium (M), and High (H) resolutions for each case.
The quantities shown are: coarsest level resolution for {\ETK} ($\delta x$), number of grid points at the edges of the coarsest
level for {\GRA} ($N_M$), 
resolution at extraction radius ($\delta x_{R}$), resolution at puncture ($\delta x_p$), total simulation 
time $t_{\rm end}$, and number of CPUs used. 
The number of points on the spherical grid where extraction is performed is 3200 for {\ETK} simulations
and 9002 for {\GRA} ($n_Q=30$).
Note in particular that the puncture resolution $\delta x_p$ matches for the corresponding resolutions
in each code. 
%
%
All physical quantities are measured in units of $M$.]}
\begin{center}
\begin{ruledtabular}
\begin{tabular}{ c c | c c c c c c} 
ID & res & $\delta x$ & $N_M$  & $\delta x_{R}$ & $\delta x_p [10^{-2}]$ & $t_{\rm end}$ & CPUs  \\
\hline
\hline
                & L & 6 & -- & 1.5 & 2.3438  & 320 &  180  \\
\etksim{42q1s0} & M & 4& -- &  1.0 & 1.5625 & 320 &  288  \\
                & H & 3 & -- & 0.75 & 1.1719 & 320 &  288  \\
\hline
                & L & 6 & -- & 1.5 & 2.3438 & 450 &  180  \\
\etksim{48q1s0} & M & 4 & -- & 1.0& 1.5625  & 450 &  288  \\
                & H & 3& --  & 0.75& 1.1719  & 450 &  288  \\
\hline
                & L & 6& --  & 1.5 & 2.3438 & 750 &  180 \\
\etksim{50q1s0} & M & 4 & -- & 1.0 & 1.5625 & 750 &  288  \\
                & H & 3 & -- & 0.75 & 1.1719  & 750 &  288  \\
\hline\hline
                & L &-- & 128  & 3.0 & 2.3438 & 500 &  768  \\
\athsim{42q1s0} & M & -- & 192 & 2.0 & 1.5625 & 500 & 1536  \\
                & H &  -- & 256 & 1.5 &1.1719 & 500 & 2560  \\
\hline
                 & L & -- & 128  & 3.0 & 2.3438  & 550 &  768  \\
\athsim{48q1s0} & M & -- & 192 & 2.0 & 1.5625  & 550 & 1536  \\
               & H & -- & 256 & 1.5 & 1.1719  & 550 & 2560  \\
\hline
                 & L & -- & 128  & 3.0 & 2.3438  & 800 &  768  \\
\athsim{50q1s0} & M & -- & 192 & 2.0 & 1.5625  & 800 & 1536  \\
               & H & -- & 256 & 1.5 & 1.1719  & 800 & 2560  \\
\end{tabular}
\end{ruledtabular}
\end{center}
\end{table}

\subsubsection{Self-convergence}
\label{sec:self-conv}

\begin{figure}[t]
   \includegraphics[width=.45\textwidth]{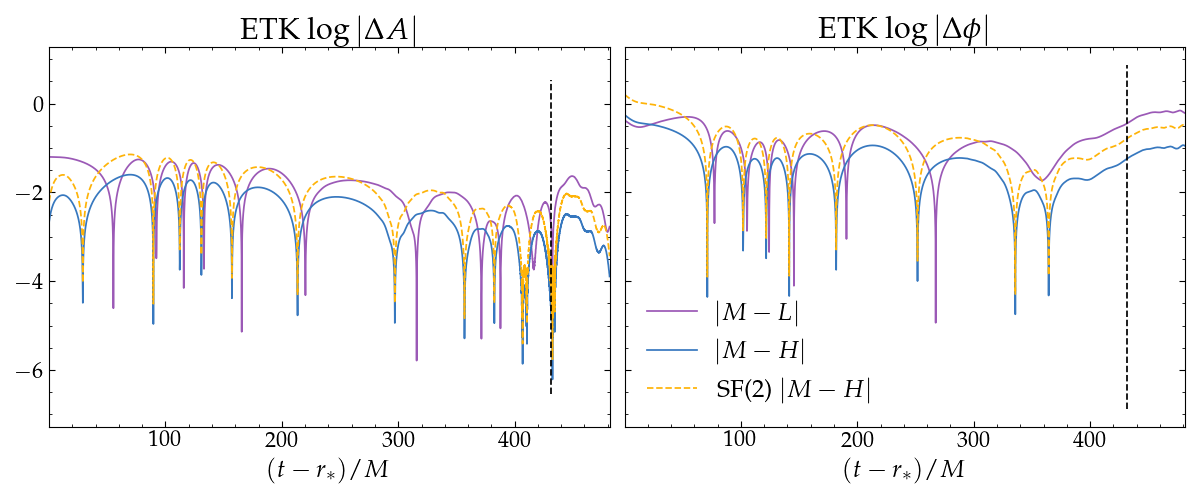}\\   	
   \includegraphics[width=.45\textwidth]{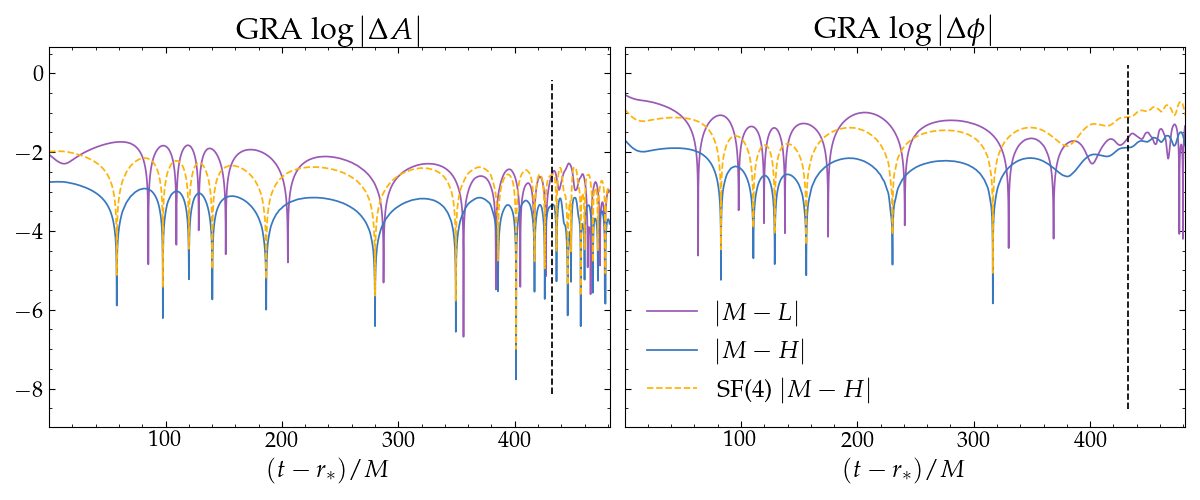}\\   
    \includegraphics[width=.45\textwidth]{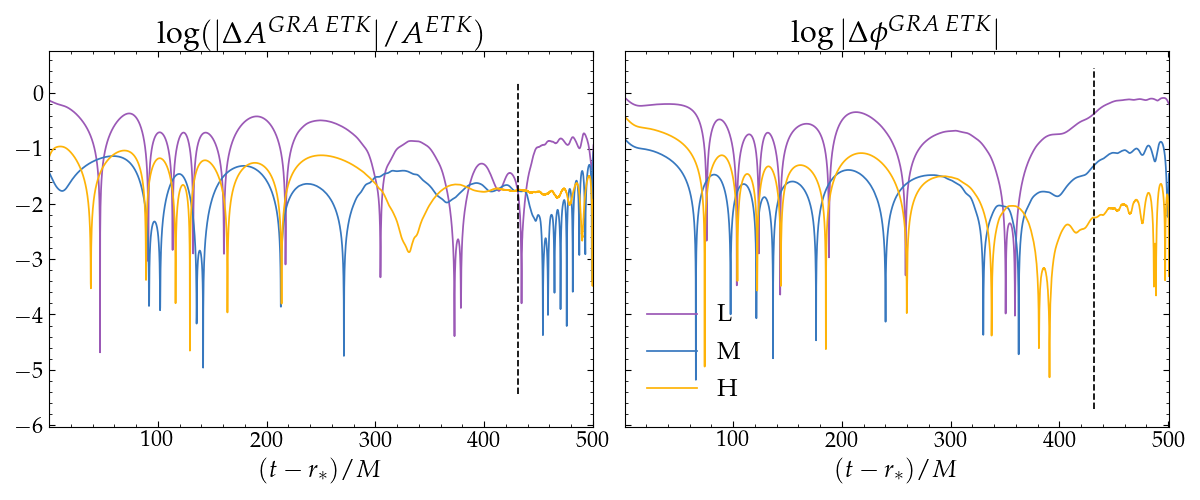}\\   
\caption{\label{fig:gra_etk}Self-convergence for  \etksim{50q1s0} (top row) and
for \athsim{50q1s0} (middle row). Displayed are the differences in amplitude and phase for
consecutive resolutions $(L-M)$, $(M-H)$. In addition, we overlay the curves for $SF(r) (M-H)$ 
which allows us to test for convergence of a given order $r$, see \eqref{eq:SF}, which we show in each panel.
Bottom row: comparison between \etksim{50q1s0} and \athsim{50q1s0} at fixed resolution. 
The simulations for $(L,M,H)$ resolutions have the same grid spacing at puncture
location, see Table~\ref{tab:self_conv_info}.
  For each case, we display the absolute value of the difference in normalized 
  amplitude and phase for every available resolution. 
}
\end{figure}

We begin by decomposing the leading modes in amplitude and phase as in \eqref{eq:strain_A_phi}.
We compare these for different resolutions as a function of time, performing 
time interpolation of third order. 
Let us denote the amplitudes and phases at a given resolution by $A_{L,M,H}$, $\phi_{L,M,H}$. 
We can claim convergence of order $r$ if 
\begin{equation}
\label{eq:conv_check}
    \frac{A_{H} - A_{M}}{A_{L}-A_{M}} \approx SF(r)
\end{equation}
\noindent where the scaling factor $SF$ is given by
\begin{equation}
\label{eq:SF}
    SF(r) = \frac{\delta x_H^r - \delta x_M^r}{\delta x_M^r - \delta x_L^r}
\end{equation}
\noindent and similarly for the phases, see e.g. \cite{Bernuzzi:2011aq}. 
Note that the left hand side of \eqref{eq:conv_check} is a varying function of time, 
so that convergence can vary during different stages of the waveform. 
We show sample results for self-convergence as a function of time for simulations \etksim{50q1s0}
and \athsim{50q1s0} in Fig. \ref{fig:gra_etk}. Note that to ease the visualization of the relation 
\eqref{eq:conv_check} we do not normalize the amplitude differences. 
We record the (normalized) amplitude and phase differences at merger, along with the waveforms 
mismatches in Table \ref{tab:self-conv-deltas}. 

Our data for \etksim{50q1s0} is compatible with second-order convergence before merger, while near 
and after merger the convergence order increases to fourth order. 
The time-dependent self-convergence results are compatible with the behaviour of the mismatches shown in 
Table \ref{tab:self-conv-deltas}, which also decrease with increasing resolution. 
In the case of \GRA{},  we observe convergence of order 4 or higher before merger, but after merger the  convergence rate for the phase worsens.
The mismatches also decrease with increasing resolution. 

The self-convergence results for the (normalized) amplitude and phase differences at merger in Table \ref{tab:self-conv-deltas} can be taken as a proxy for the accuracy of the NR simulations when comparing to EOB. Roughly speaking,  differences between Medium resolution simulations (used for the bulk of our simulations) and High resolution is around $0.1 \%$ for the amplitude and of the order of $0.01$ radians for the phases.

%

\begin{table}[h]
\caption{\label{tab:self-conv-deltas}Self-convergence results for the $\ell=m=2$ strain mode for 
consecutive resolutions for {\ETK} and {\GRA} at merger time. From left to right, the columns 
report: the relative amplitude differences, the phase differences and the unfaithfulness
between two consecutive resolutions.}
\begin{center}
\begin{ruledtabular}
\begin{tabular}{ c c | c c c} 
ID & res$_{1,2}$ & $\Delta A^{{\rm res}_1-{\rm res}_2}_{\rm 22, mrg}/A^{{\rm res}_1}_{\rm 22, mrg} $ & $\Delta \phi^{{\rm res}_1-{\rm res}_2}_{\rm 22, mrg}$ & $\mismatch$ \\ 
 \hline 
 \hline 
\etksim{42q1s0} &L,M & $ -1.86  \cdot 10^{-3} $ & $ 2.265  \cdot 10^{-3} $ & $ 7.469  \cdot 10^{-6} $ \\ 
\etksim{42q1s0} &M,H & $ -3.29  \cdot 10^{-3} $ & $ -2.95  \cdot 10^{-3} $ & $ 1.101  \cdot 10^{-4} $ \\ 
\hline 
\etksim{48q1s0} &L,M & $ -9.73  \cdot 10^{-3} $ & $ -4.55  \cdot 10^{-2} $ & $ 2.048  \cdot 10^{-4} $ \\ 
\etksim{48q1s0} &MH & $ -3.47  \cdot 10^{-3} $ & $ -2.02  \cdot 10^{-2} $ & $ 7.393  \cdot 10^{-5} $ \\ 
\hline 
\etksim{50q1s0} &L,M & $ -1.02  \cdot 10^{-3} $ & $ -3.63  \cdot 10^{-1} $ & $ 1.917  \cdot 10^{-3} $ \\ 
\etksim{50q1s0} &M,H & $ -1.33  \cdot 10^{-5} $ & $ -5.85  \cdot 10^{-2} $ & $ 5.463  \cdot 10^{-5} $ \\
\hline
\hline
\athsim{42q1s0} &L,M & $ -1.32  \cdot 10^{-3} $ & $ 5.257  \cdot 10^{-3} $ & $ 9.756  \cdot 10^{-6} $ \\ 
\athsim{42q1s0} &M,H & $ -1.14  \cdot 10^{-3} $ & $ 1.715  \cdot 10^{-3} $ & $ 1.332  \cdot 10^{-6} $ \\ 
\hline 
\athsim{48q1s0} &L,M & $ -2.74  \cdot 10^{-3} $ & $ 1.686  \cdot 10^{-2} $ & $ 1.352  \cdot 10^{-5} $ \\ 
\athsim{48q1s0} &M,H & $ -2.65  \cdot 10^{-4} $ & $ -8.38  \cdot 10^{-3} $ & $ 9.574  \cdot 10^{-6} $ \\ 
\hline 
\athsim{50q1s0} &L,M & $ -6.56  \cdot 10^{-3} $ & $ 2.204  \cdot 10^{-2} $ & $ 9.405  \cdot 10^{-5} $ \\ 
\athsim{50q1s0} &M,H & $ -1.06  \cdot 10^{-3} $ & $ -1.30  \cdot 10^{-2} $ & $ 1.484  \cdot 10^{-5} $
\end{tabular}
\end{ruledtabular}
\end{center}
\end{table}

\subsubsection{Code comparison}
\label{sec:code_cf}

We now focus on the comparison between \ETK{} and \GRA{}. We display the amplitude 
and phase differences as a function of time for the leading modes of \etksim{50q1s0} and \athsim{50q1s0} at the bottom of  Fig. \ref{fig:gra_etk}. The
cases \etksim{42q1s0, 48q1s0} and \athsim{42q1s0, 48q1s0} behave similarly. 
We summarize the information regarding the differences at merger and mismatches in Table \ref{tab:code_comparison}. 
For the three simulations, we find that the waveforms are most coincident at Medium resolution. 
In this case, the amplitude difference at merger is of order $2 \%$, being always higher for 
\GRA. This is compatible with the results of \cite{Daszuta:2021ecf} which found a 
$2 \%$ difference at merger with the \BAM{} code \cite{Brugmann:2008zz}. 
While we observe some decrease of the amplitude difference at merger values for High resolution, 
it is not clear whether they  will converge away with increasing resolution further.  
On the other hand, the phases at merger appear to be converging with increasing resolution, 
the differences ranging from $10^{-2}$ to $10^{-3}$. 
At Medium resolution, which is the one used for {\ETK} in the bulk of this work, we find that the differences with \GRA{} are roughly of the order of $1 \%$ for the amplitude and $0.01$ radians for the phase. 

The mismatches for {\ETK} and {\GRA} leading modes at Medium resolution are of order 
$10^{-5}$, and show convergent behaviour for \etksim{48q1s0, 50q1s0} (\athsim{48q1s0, 50q1s0}) but not for \etksim{42q1s0} (\athsim{42q1s0}). 

We emphasize that the comparisons discussed above involve the leading modes of the 
extrapolated strain. However, we should keep in mind that the NR results are also 
affected by our choice of time integration and extrapolation to infinity. We provide a comparison 
of the raw data by considering the unextrapolated $\psi_4$ scalars produced by {\ETK} 
and {\GRA} in Appendix~\ref{app:cc_psi4}. 
\begin{table}[h]
\caption{\label{tab:code_comparison}Comparing the $\ell=m=2$ strain waveform at merger obtained with {\ETK} and {\GRA}.
From left to right, the columns report: the relative amplitude differences, the phase differences at merger and the unfaithfulness
for the three configurations simulated with both codes at (L,M,H) resolution.}
\begin{center}
\begin{ruledtabular}
\begin{tabular}{ c c | c c c } 

ID & res & $\Delta A^{\rm GRAETK}_{\rm 22, mrg} / A^{GRA}_{\rm 22, mrg}$ & $\Delta \phi^{\rm GRAETK}_{\rm 22,mrg}$ & $\mismatch$ \\ 
 \hline 
 \hline 
 &L & $ 1.640  \cdot 10^{-2} $ & $ 1.731  \cdot 10^{-2} $ & $ 5.417  \cdot 10^{-5} $ \\ 
{\tt42q1s0} &M & $ 1.567  \cdot 10^{-2} $ & $ 1.364  \cdot 10^{-2} $ & $ 1.788  \cdot 10^{-5} $ \\ 
 &H & $ 1.355  \cdot 10^{-2} $ & $ 9.073  \cdot 10^{-3} $ & $ 9.145  \cdot 10^{-5} $ \\ 
\hline 
 & L & $ 2.427  \cdot 10^{-2} $ & $ 8.193  \cdot 10^{-2} $ & $ 2.216  \cdot 10^{-4} $ \\ 
{\tt 48q1s0} & M & $ 1.704  \cdot 10^{-2} $ & $ 1.973  \cdot 10^{-2} $ & $ 1.991  \cdot 10^{-5} $ \\ 
 & H & $ 1.388  \cdot 10^{-2} $ & $ 7.851  \cdot 10^{-3} $ & $ 1.538  \cdot 10^{-5} $ \\ 
\hline 
 & L & $ 1.099  \cdot 10^{-2} $ & $ 4.363  \cdot 10^{-1} $ & $ 2.117  \cdot 10^{-3} $ \\ 
{\tt 50q1s0} & M & $ 1.737  \cdot 10^{-2} $ & $ 5.084  \cdot 10^{-2} $ & $ 2.721  \cdot 10^{-5} $ \\ 
 & H & $ 1.848  \cdot 10^{-2} $ & $ 5.683  \cdot 10^{-3} $ & $ 2.206  \cdot 10^{-6} $ 
\end{tabular}
\end{ruledtabular}
\end{center}
\end{table}

\section{Effective-One-Body model}
\label{sec:EOB}
In this work we will use the eccentric version of the \TEOB{}~\cite{Nagar:2020pcj,Riemenschneider:2021ppj} 
effective-one-body (EOB)-based~\cite{Buonanno:1998gg,Buonanno:2005xu} waveform model, 
dubbed \TEOB{-Dalì}, as defined in Refs.~\cite{Nagar:2021gss,Bonino:2022hkj}. The promotion of the quasi-circular model
to the eccentric case follows the idea of Ref.~\cite{Chiaramello:2020ehz} of incorporating 
non-circular effects in the Newtonian prefactors in both the waveform and radiation 
reaction\footnote{Higher order 
eccentricity-dependent PN terms have been computed in Refs.~\cite{Khalil:2021txt, 
Placidi:2021rkh, Albanesi:2022xge}, but we will neglect them for the purpose of this work.}.
We refer the reader to Ref.~\cite{Nagar:2021gss} and references therein for most of 
the technical details of the model. Here it is only worth recalling that the EOB description
of the merger and ringdown is based on suitable fits of {\it quasi-circular} ringdown 
waveforms~\cite{Damour:2014yha,Nagar:2020pcj}. This approximation looks sufficiently
accurate for bound configurations, because the eccentric inspiral has the time to progressively 
circularize towards merger~\cite{Nagar:2021gss}. By contrast, for dynamical capture configurations
this is not the case and the quasi-circular ringdown can be inaccurate~\cite{Albanesi:2021rby,Gamba:2021gap}.
For this reason, one of the final goals of this paper is to show how to use our new
NR simulations to suitably improve the model during merger and ringdown.
Before discussing this, let us recall that any EOB model essentially depends on 
two sets of parameters that are informed by NR simulations: (i) one the one 
hand there are those that directly appear in the dynamics, i.e. as effective modifications
to the EOB Hamiltonian. Belonging to this class are the effective 5PN parameter 
$a_6^c$ entering the $A$ potential of \TEOB or the effective next-to-next-to-next-to-leading 
order parameter $c_3$ used to improve the corresponding spin-orbit coupling); (ii) on the other
hand, there are parameters used to improve the shape of the waveform during plunge up to 
merger via the next-to-quasi-circular (NQC) correction or to accurately describe the 
postmerger-ringdown signal. In this paper we do not focus on exploring the effect of
the dynamical parameters, the values of which we set to those determined in~\cite{Nagar:2021gss},
but rather explore the impact of informing the remaining ones using data from our dynamical
capture BBH simulations. In particular, we want to: (i) obtain new, NR-informed, 
next-to-quasi-circular corrections to the waveform and (ii)use a new NR-informed 
merger and ringdown model. 
Since our NR simulations cover parameter space in a sparse way, we can 
only use the NR information separately for each datasets, and cannot present 
global fits of the NR-informed parameters designed to cover the full parameter space. 
We therefore aim to illustrate what is needed to improve the current model for hyperbolic 
capture and quantify the improvement, leaving the development of global fits for 
future work.
%
From now on, we will focus only on the quadrupole waveform for simplicity, although
the same approach could be extended to the other multipoles.
We decompose it in amplitude and phase as
\be
h_{22}=A_{22} e^{-i\phi_{22}} \ , 
\ee
where both $(A_{22},\phi_{22})$ are functions of time $t$. 
The waveform frequency is $\omega_{22}\equiv \dot{\phi}_{22}$. In addition,
we will use the $\nu$-normalized waveform amplitude $\hat{A}_{22}\equiv A_{22}/\nu$.
Following the standard \TEOB{} procedure for quasi-circular binaries~\cite{Nagar:2020pcj,Riemenschneider:2021ppj},
we count four parameters related to NQC corrections and seven parameters
needed to model the postmerger-ringdown part of the waveform. In practice, 
one needs to extract  {\it eleven} numbers from each NR simulation.
The NR-informed parameters are:
\begin{itemize}
 \item[(i)]
 The four NQC parameters $(a^{22}_1,a^{22}_2,b^{22}_1,b^{22}_2)$ that enter the NQC waveform correction
 \be
 \hat{h}_{22}^{\rm NQC}=(1 + a_1^{22} n^{22}_1 + a_2^{22} n^{22}_2)e^{{\rm i}(b_1 n^{22}_3 + b_2 n^{22}_4)} \ .
 \ee
 Here, the functions $n_i^{22}$ define the NQC basis (see in particular text 
 around Eq.~(13) of Ref.~\cite{Nagar:2021gss}),
 are determined by imposing continuity between the EOB waveform and the NR amplitude
 and frequency values $(\hat{A}^{\rm NR}_{\rm NQC}, \dot{\hat{A}}^{\rm NR}_{\rm NQC}, \omega^{\rm NR}_{\rm NQC},\dot{\omega}^{\rm NR}_{\rm NQC})$ 
 evaluated at the NQC extraction time $t_{\rm NQC}^{\rm NR}$. For each mode,  $t_{\rm NQC}^{\rm NR} = t^{\rm NR}_{\rm A^{\rm max}} + 2 M$, 
 where $t^{\rm NR}_{\rm A^{\rm max}}$ is the location of the amplitude's peak.    
 \item[(ii)] The five parameters entering the ringdown template as introduced in Ref.~\cite{Damour:2014yha}
    \begin{equation}
     \theta^{\rm RD} = (\hat{A}_{\rm max}, \omega_{\rm max}, c_{3}^A, c^3_\phi, c^4_\phi )\, .
    \end{equation} 
    The values of the amplitude and frequency at amplitude maximum $(A_{\rm max},\omega_{\rm max})$ are directly
    read from the NR waveform. By contrast, the parameters $(c_{3}^A, c^3_\phi, c^4_\phi)$ are obtained
    by fitting the postmerger-ringdown waveform following~\cite{Damour:2014yha}.
 \item[(iii)]
 The postmerger template also depends on the  (complex) frequency of the first two quasi-normal-modes that
 are determined from the mass and dimensionless spin of the final black hole $(M_f,\hat{a}_f)$ interpolating the 
 tables of Ref.~\cite{Berti:2005ys}. 
 \end{itemize}
Table~\ref{tab:eob_parameters} lists all the NR-informed parameter for each NR dataset considered. The last two
columns also report the EOB/NR difference in the binding energy at merger and the phase difference 
$\Delta \phi^{\rm EOBNR}_{22}\equiv \phi^{\rm EOB}_{22}-\phi^{\rm NR}_{22}$. Note that the latter is computed,
as we will see below, using an {\it improved} model with the NR-informed ringdown for each dataset. The
origin of these numbers will be discussed in detail in Sec.~\ref{sec:eobnr} below.
\begin{figure}[h]
\includegraphics[width=.49\textwidth]{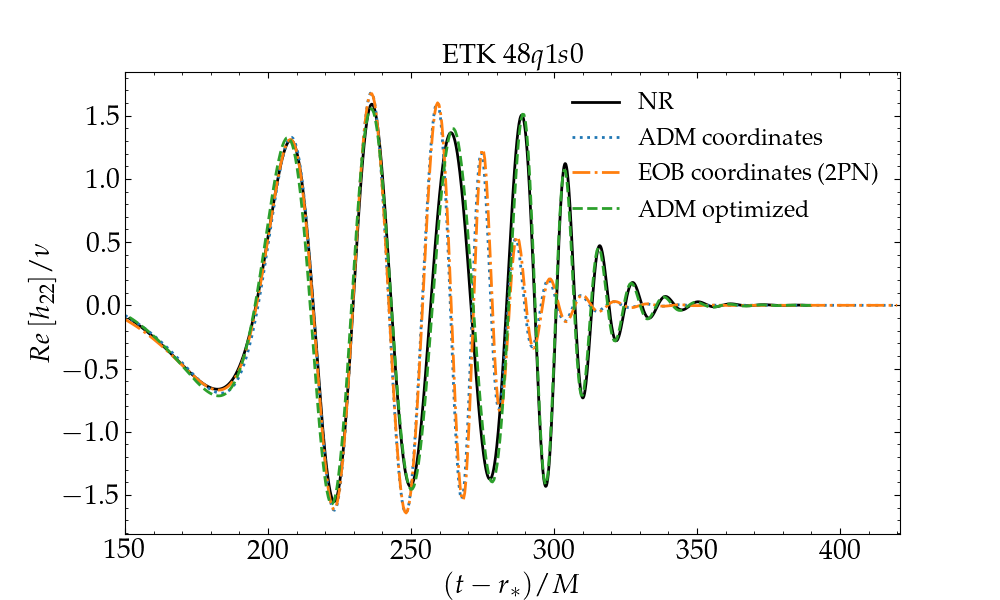}
\vspace{5mm}
\includegraphics[width=.49\textwidth]{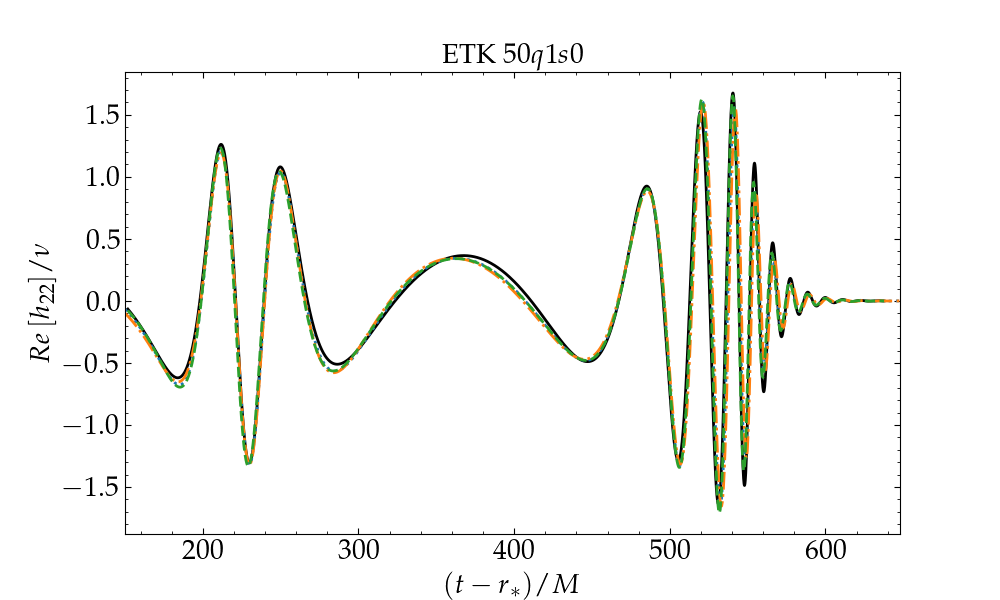}
\caption{Relation between NR initial data and EOB ones for two selected configurations. The procedure 
of tuning phenomenological corrections to $(E_0^{\rm ADM},J_0^{\rm ADM})$ using 
Eqs.~\eqref{eq:E0adm}-\eqref{eq:J0adm} is globally more efficient (especially for the \etksim{48q1s0} dataset) 
than the straightforward transformation from ADM to EOB coordinates.} 
\label{fig:2PN_test} 
\end{figure}

\begin{table*}[t] 
\caption{\label{tab:eob_parameters}Informing the EOB model with NR data. From left to right the columns report: 
the name of the dataset, the $\nu$-normalized quadrupole waveform amplitude at merger time $\hat{A}_{22}^{\rm max}=A_{22}^{\rm max}/\nu$
and the corresponding frequency; the mass $M_f$ and dimensionless angular momentum $\hat{a}_f\equiv J_f/M_f^2$ of the final remnant 
black hole; the NR values used to determine the NQC corrections; the NR-informed parameters of the postmerger fit. The last two columns
report the EOBNR difference in the binding energy per reduced mass at merger $\delta E_{\rm b,\,mrg}^{\rm EOBNR}\equiv E_{\rm b, mrg}^{\rm EOB}-E_{\rm b, mrg}^{\rm NR}$
and the corresponding phase difference at merger obtained using the NR-informed upgraded model.}
\begin{center}
\begin{ruledtabular}
\begin{tabular}{ c c | c c c c c c c c c c c| c c} 
\# & ID  &  $\hat{A}^{\rm max}_{22}$ & $\omega^{\rm max}_{22}$ & $M_f$ & $a_f$ & $ \hat{A}^{\rm NQC}_{\rm NR}$ & $ \dot {\hat{A}}_{\rm NR}^{\rm NQC}$ & $\omega^{\rm NQC}_{\rm NR}$ & $\dot \omega^{\rm NQC}_{\rm NR}$ & $c_3^A$ & $c_3^\phi$ & $c_4^\phi$  &  $\tilde \delta E_{\rm b, mrg} $ &  $\Delta \phi^{\rm EOBNR}_{22, \rm mrg}$[rad]\\ 
\hline 
 \hline 
1 &\etksim{37q1s0} & $ 1.419 $ &  $ 0.335 $ &  $ 0.978 $ &  $ 0.685 $ &  $ 1.399 $ &  $ -0.020 $ &  $ 0.369 $ &  $ 0.017 $ &  $ -0.016 $ &  $ 1.74076 $ &  $ 0.39046 $ &  $+ 0.0045 $ &  $ -0.003 $ \\ 
2 &\etksim{42q1s0} & $ 1.675 $ &  $ 0.370 $ &  $ 0.965 $ &  $ 0.725 $ &  $ 1.659 $ &  $ -0.016 $ &  $ 0.398 $ &  $ 0.014 $ &  $ -0.187 $ &  $ 2.48789 $ &  $ 0.73609 $ &  $+ 0.0038 $ &  $ -0.008 $ \\ 
3 & \etksim{44q1s0} & $ 1.768 $ &  $ 0.388 $ &  $ 0.958 $ &  $ 0.728 $ &  $ 1.754 $ &  $ -0.015 $ &  $ 0.410 $ &  $ 0.011 $ &  $ -0.243 $ &  $ 3.24197 $ &  $ 1.19834 $ &  $+ 0.0039 $ &  $ -0.029 $ \\ 
4 & \etksim{46q1s0} & $ 1.821 $ &  $ 0.377 $ &  $ 0.949 $ &  $ 0.716 $ &  $ 1.814 $ &  $ -0.007 $ &  $ 0.394 $ &  $ 0.009 $ &  $ -0.510 $ &  $ 5.30479 $ &  $ 1.33265 $ &  $+ 0.0025 $ &  $ -0.015 $ \\ 
5 &\etksim{48q1s0} & $ 1.518 $ &  $ 0.346 $ &  $ 0.950 $ &  $ 0.675 $ &  $ 1.509 $ &  $ -0.010 $ &  $ 0.370 $ &  $ 0.012 $ &  $ -0.422 $ &  $ 4.42020 $ &  $ 1.91359 $ &  $+ 0.0030 $ &  $ -0.096 $ \\ 
6 &\etksim{50q1s0} & $ 1.679 $ &  $ 0.364 $ &  $ 0.949 $ &  $ 0.697 $ &  $ 1.671 $ &  $ -0.008 $ &  $ 0.383 $ &  $ 0.010 $ &  $ -0.477 $ &  $ 4.80545 $ &  $ 1.26372 $ &  $+ 0.0033 $ &  $ -0.011 $ \\ 
7 &\etksim{42q1s050--} & $ 1.429 $ &  $ 0.307 $ &  $ 0.981 $ &  $ 0.515 $ &  $ 1.409 $ &  $ -0.020 $ &  $ 0.336 $ &  $ 0.014 $ &  $ 0.001 $ &  $ 1.59410 $ &  $ 0.28785 $ &  $+ 0.0041 $ &  $ -0.030 $ \\ 
8 &\etksim{42q1s025--} & $ 1.522 $ &  $ 0.338 $ &  $ 0.976 $ &  $ 0.626 $ &  $ 1.506 $ &  $ -0.017 $ &  $ 0.367 $ &  $ 0.014 $ &  $ -0.072 $ &  $ 2.07120 $ &  $ 0.53770 $ &  $+ 0.0040 $ &  $ -0.012 $ \\ 
9 &\etksim{42q1s025++} & $ 1.863 $ &  $ 0.403 $ &  $ 0.943 $ &  $ 0.793 $ &  $ 1.849 $ &  $ -0.015 $ &  $ 0.425 $ &  $ 0.012 $ &  $ -0.459 $ &  $ 5.39366 $ &  $ 2.68358 $ &  $+ 0.0025 $ &  $ -0.035 $ \\ 
10 &\etksim{42q1s050++} & $ 1.487 $ &  $ 0.428 $ &  $ 0.940 $ &  $ 0.832 $ &  $ 1.476 $ &  $ -0.012 $ &  $ 0.453 $ &  $ 0.013 $ &  $ -0.093 $ &  $ 2.87101 $ &  $ 1.14302 $ &  $+ 0.0057 $ &  $ 0.014 $ \\ 
11 &\etksim{42q1s050+-} & $ 1.676 $ &  $ 0.370 $ &  $ 0.965 $ &  $ 0.724 $ &  $ 1.660 $ &  $ -0.017 $ &  $ 0.398 $ &  $ 0.014 $ &  $ -0.181 $ &  $ 2.49269 $ &  $ 0.73682 $ &  $+ 0.0038 $ &  $ -0.012 $ \\ 
\hline
12 &\etksim{42q1.5s0} & $ 1.763 $ &  $ 0.374 $ &  $ 0.962 $ &  $ 0.707 $ &  $ 1.748 $ &  $ -0.015 $ &  $ 0.398 $ &  $ 0.012 $ &  $ -0.297 $ &  $ 3.17798 $ &  $ 1.12475 $ &  $+ 0.0029 $ &  $- 0.003 $ \\ 
13 &\etksim{42q2s0} & $ 1.734 $ &  $ 0.327 $ &  $ 0.957 $ &  $ 0.633 $ &  $ 1.730 $ &  $ -0.004 $ &  $ 0.338 $ &  $ 0.006 $ &  $ -0.845 $ &  $ 9.62619 $ &  $ 1.39233 $ &  $ -0.0003 $ &  $ -0.026 $ \\ 
14 &\etksim{42q2.15s0} & $ 1.681 $ &  $ 0.337 $ &  $ 0.963 $ &  $ 0.630 $ &  $ 1.674 $ &  $ -0.007 $ &  $ 0.358 $ &  $ 0.011 $ &  $ -0.597 $ &  $ -0.00004 $ &  $ 0.00001 $ &  $+ 0.0030 $ &  $+ 0.032 $ \\
\end{tabular}
\end{ruledtabular}
\end{center}
\end{table*}

\begin{figure*}[t]
\includegraphics[width=.4\textwidth]{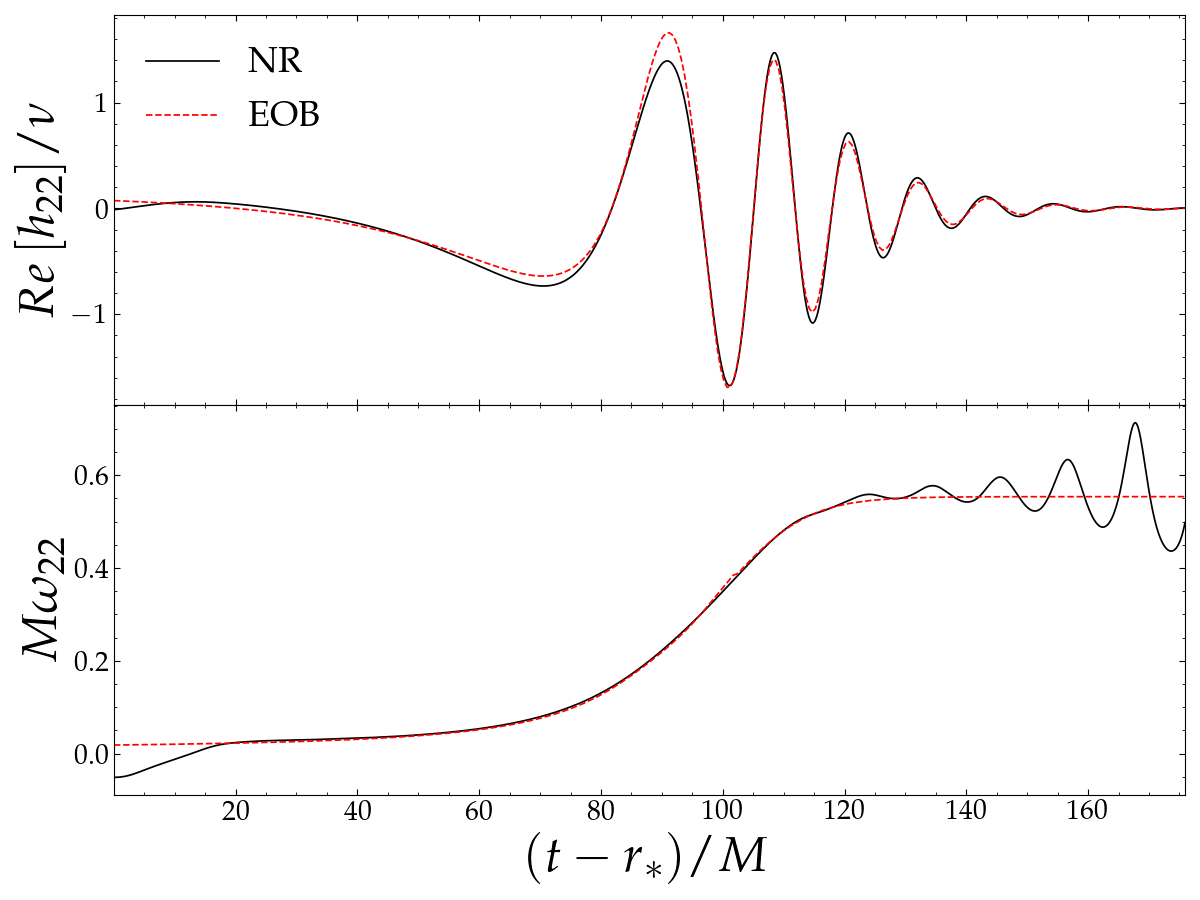}
\hspace{5mm}
\includegraphics[width=.4\textwidth]{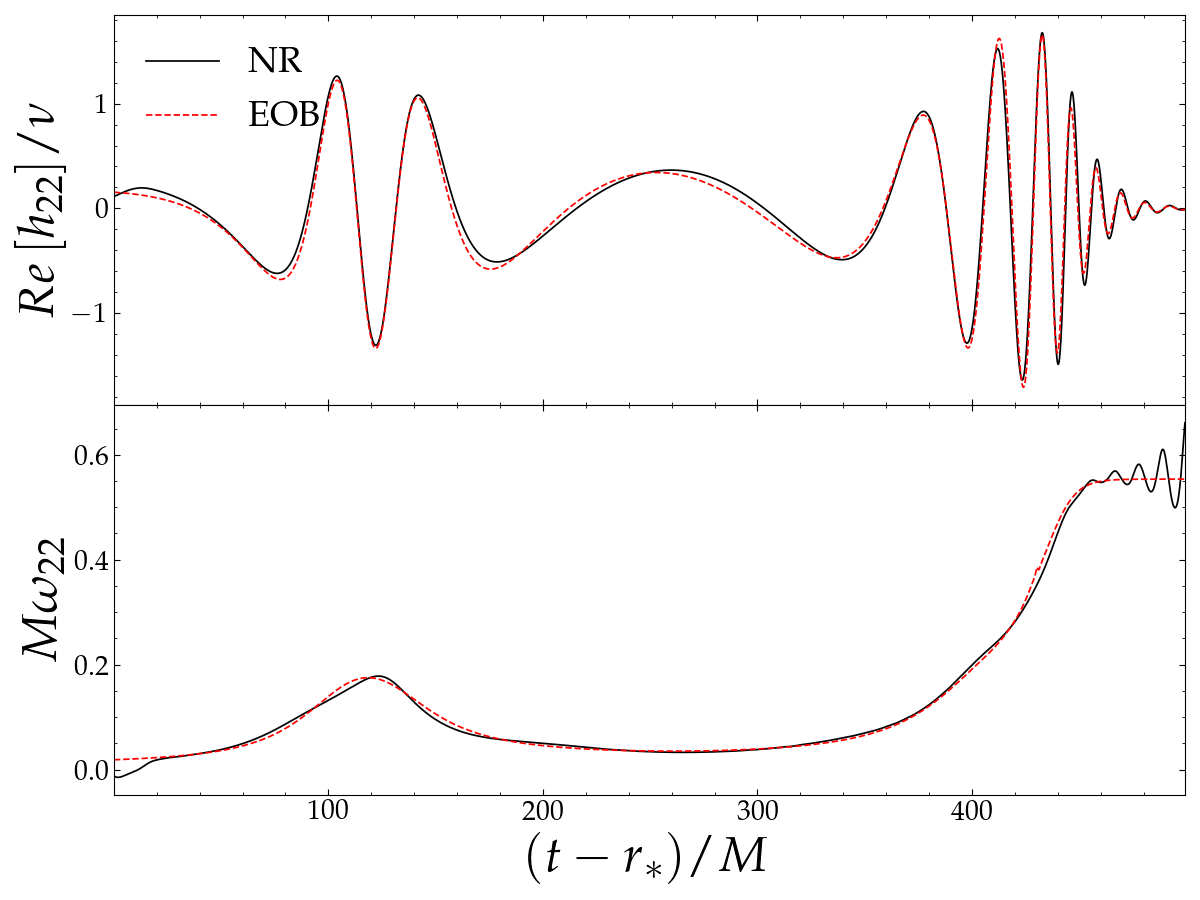}\\
\vspace{6mm}
\includegraphics[width=.4\textwidth]{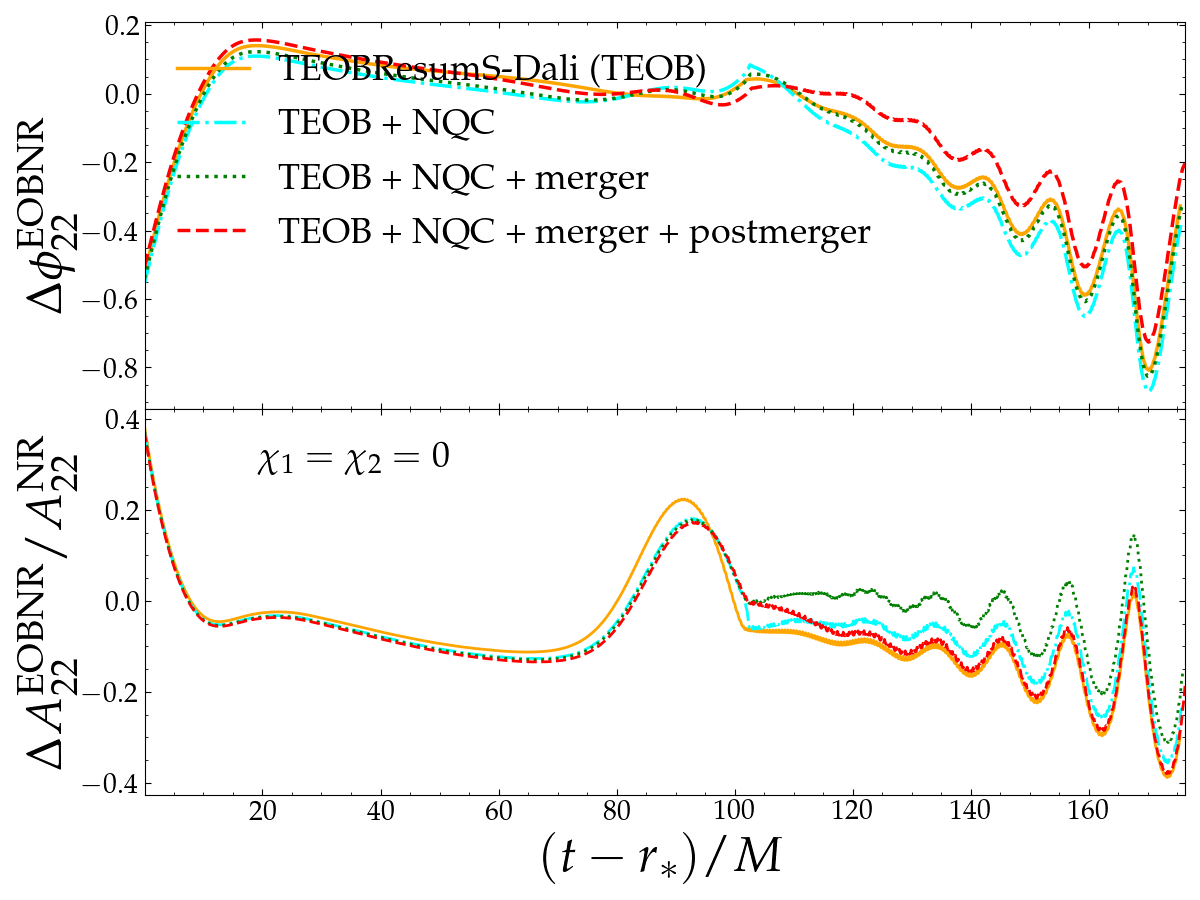}
\hspace{5mm}
\includegraphics[width=.4\textwidth]{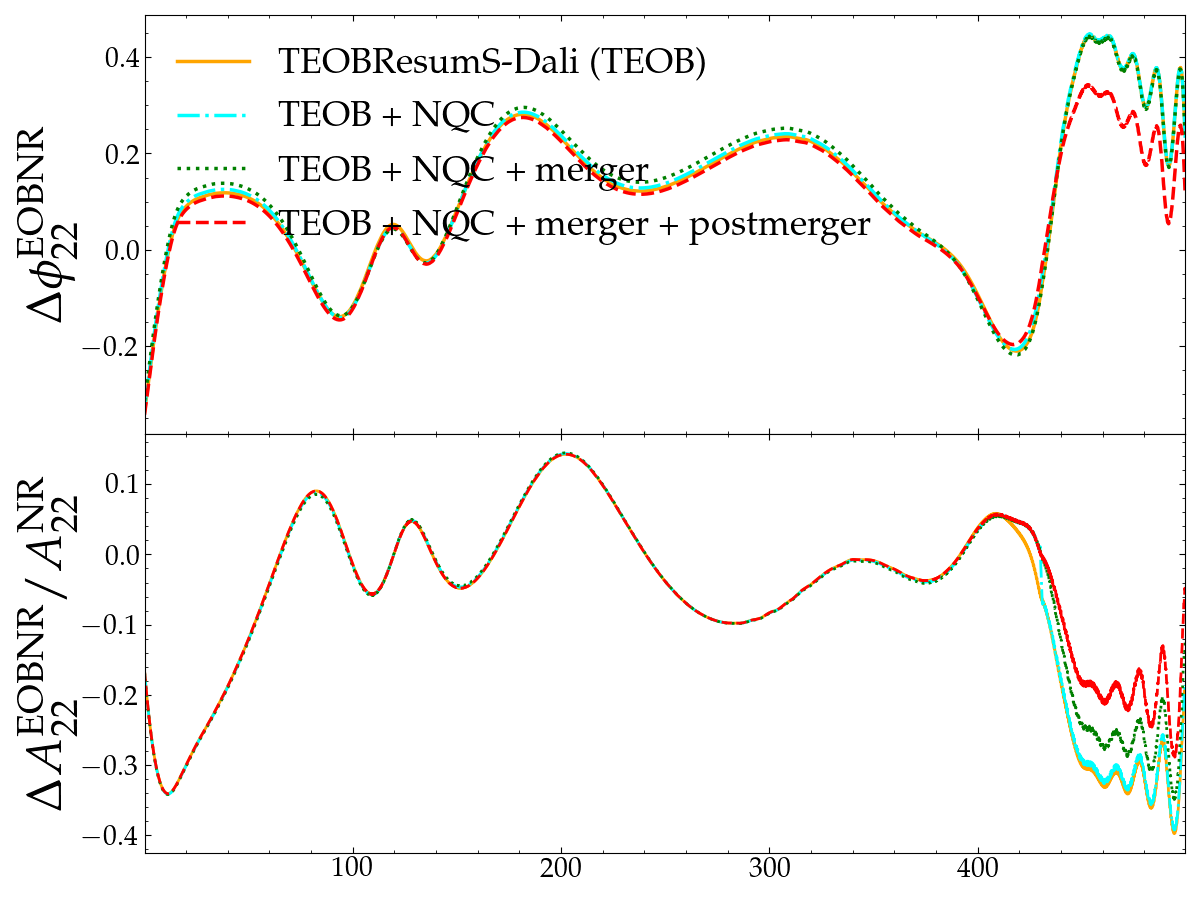}\\
\caption{Equal-mass, nonspinning case. EOB/NR waveform comparison.
Top panels: $\ell=m=2$ real part and frequency evolution for \etksim{42q1s0} (left) and \etksim{50q1s0} (right). 
Bottom panels: EOB/NR phase and (relative) amplitude differences obtained by increasing progressively the
amount of NR input to describe the merger and ringdown part.} 
\label{fig:nr_eob_42_50_all} 
\end{figure*}

\subsection{Connecting NR initial data with EOB ones}
\label{sec:initial_data}

In order to compare our results from NR and EOB, we need to specify which parameters shall 
we input in \TEOB{} based on the initial NR data. Initial data for dynamical capture in 
\TEOB{} are given by fixing the mass ratio $q$, initial energy $E_{\rm EOB}/M$, dimensionless initial orbital angular 
momentum $p_\varphi=L^{\rm EOB}/(\mu M)$, dimensionless spins $\chi_{1,2}^{\rm EOB}\equiv S_i/m_i^2$
and initial separation $r_{\rm EOB}$. In practice, the initial EOB parameters 
are obtained by identifying the NR and EOB spin values, i.e. $\chi_{1,2}^{\rm EOB}=\chi_{1,2}^{\rm NR}$, 
and matching the other dynamical quantities to the initial ADM values from NR simulations as
\begin{align}
\label{eq:E0adm}
    E_{\rm EOB} & = E_0^{\rm ADM} + \Delta E_{\rm EOB} \ , \\
\label{eq:J0adm}
    L_{\rm EOB} & = J_0^{\rm ADM} - (S_1 + S_2) + \Delta J_{\rm EOB} \ , \\
    r_{\rm EOB} & = D + \Delta r_{\rm EOB} \ ,
\end{align}
where $(\Delta E_{\rm EOB},\Delta J_{\rm EOB},\Delta r_{\rm EOB})$ are arbitrary corrections
to the NR quantities to be determined as follows.
We choose the initial EOB distance $\Delta r_{\rm EOB}$ such that the EOB waveforms extend 
to earlier times at least as much as the NR simulations. For this reason, we choose $\Delta r_{\rm EOB} = 4 M$.  
Following \cite{Gamba:2021gap}, values of $(\Delta E_{\rm EOB},\Delta J_{\rm EOB})$ 
are chosen by minimizing the EOB/NR mismatch. More precisely, when performing this minimization 
we consider both the ring-down and NQC information in the EOB waveform. We implement this using the 
algorithm \texttt{dual{\_}annealing} from the \texttt{scipy} Python library \cite{Virtanen:2020}. 
We find that we obtain a good match for the signal length considering $|\Delta E_{\rm EOB}| < 0.0007 E_{\rm ADM}$,
$|\Delta J_{\rm EOB}| < 0.006 J_{\rm ADM}$, see Table~\ref{tab:ETK_sims}. 

Although this method is efficient and successful, we have to remind the reader that at a rigorous
mathematical level the initial puncture parameters expressed in ADM coordinates (relative separation
and momenta) should be connected to the corresponding EOB ones by using the corresponding canonical
transformation. This was originally obtained in Ref.~\cite{Buonanno:1998gg} at 2PN accuracy. One of the
use of this transformation, already suggested in Ref.~\cite{Buonanno:1998gg} was to provide small-eccentricity
initial data for NR simulations. This idea was eventually implemented in Ref.~\cite{Walther:2009ng} at
3PN accuracy. For completeness, here we also explored this route by converting  the ADM quantities 
$(\vec P^{\rm ADM}, D)$ to EOB ones $(E_{\rm EOB}, p_\varphi^{\rm EOB}, r_{\rm EOB})$ using the 
canonical transformation at 2PN.
Figure~\ref{fig:2PN_test} focuses on two datasets, \etksim{48q1s0} and \etksim{50q1s0}:
the NR waveforms (black) are compared with EOB waveforms obtained with different choices
of initial conditions.  We note that the impact of the optimization and of the canonical transformation
(that typically is rather small) depend on the configuration. For \etksim{48q1s0} the effect of the
ADM/EOB canonical transformation is practically negligible, while the phenomenological optimization
is more efficient in obtaining the correct initial data. By contrast, for \etksim{50q1s0} the various
choices are practically equivalent.

\begin{figure*}[t]
    \includegraphics[width=.3\textwidth]{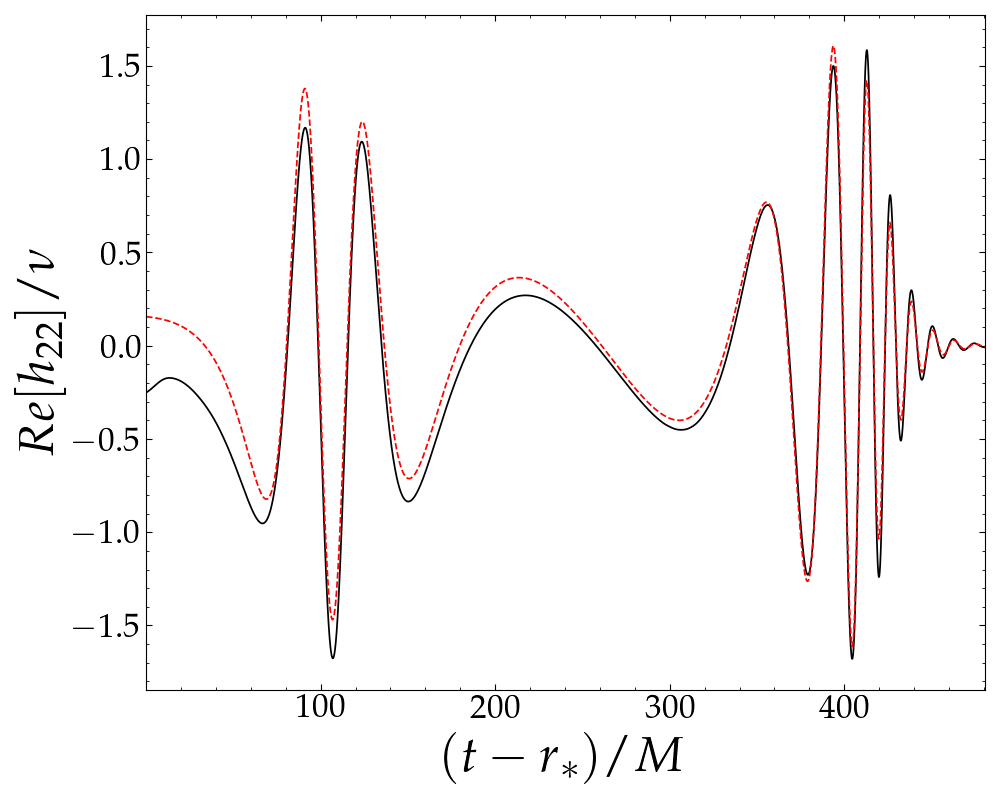}
     \includegraphics[width=.3\textwidth]{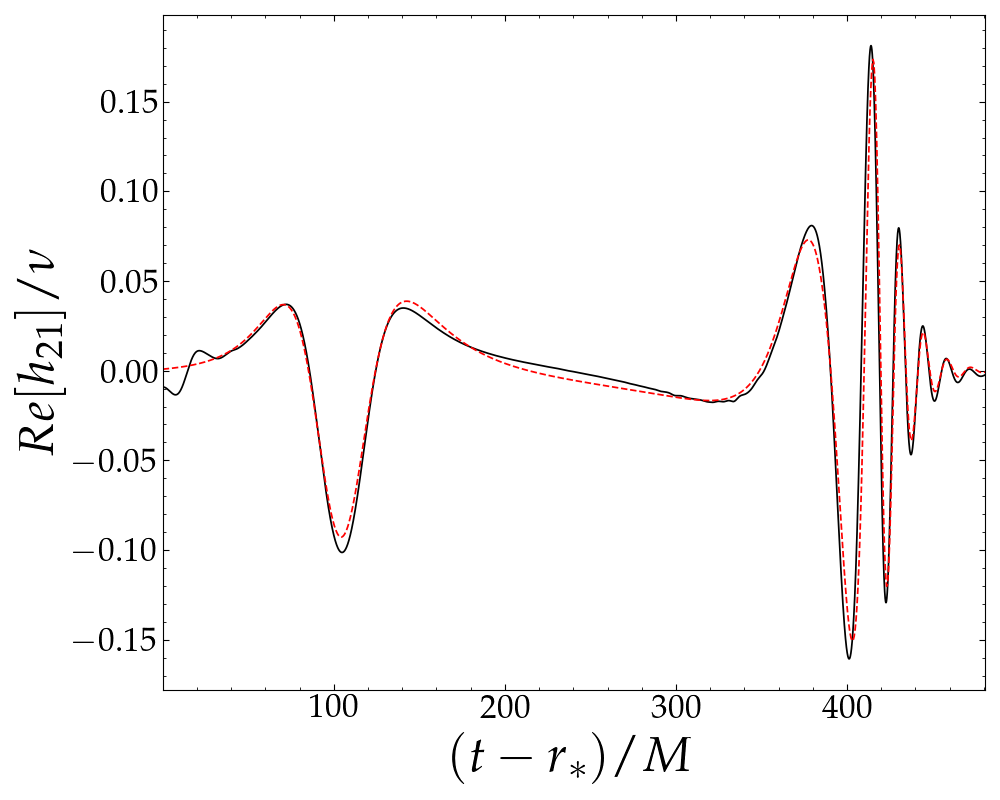}
    \includegraphics[width=.3\textwidth]{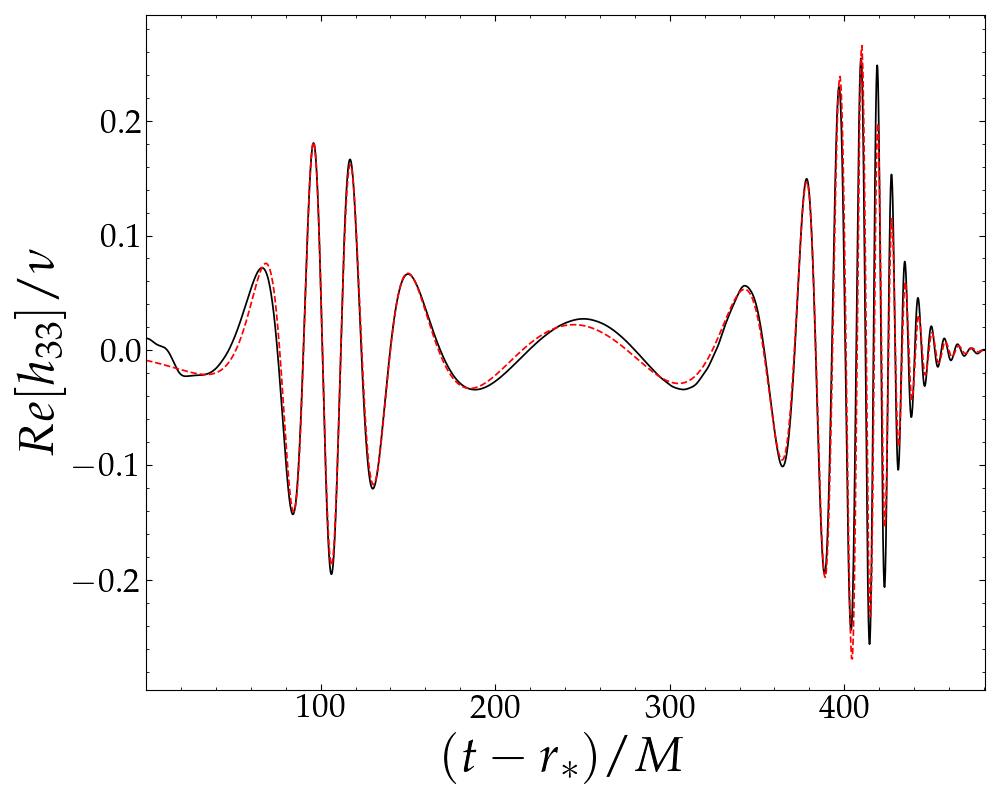}\\
    \vspace{5mm}
    \includegraphics[width=.3\textwidth]{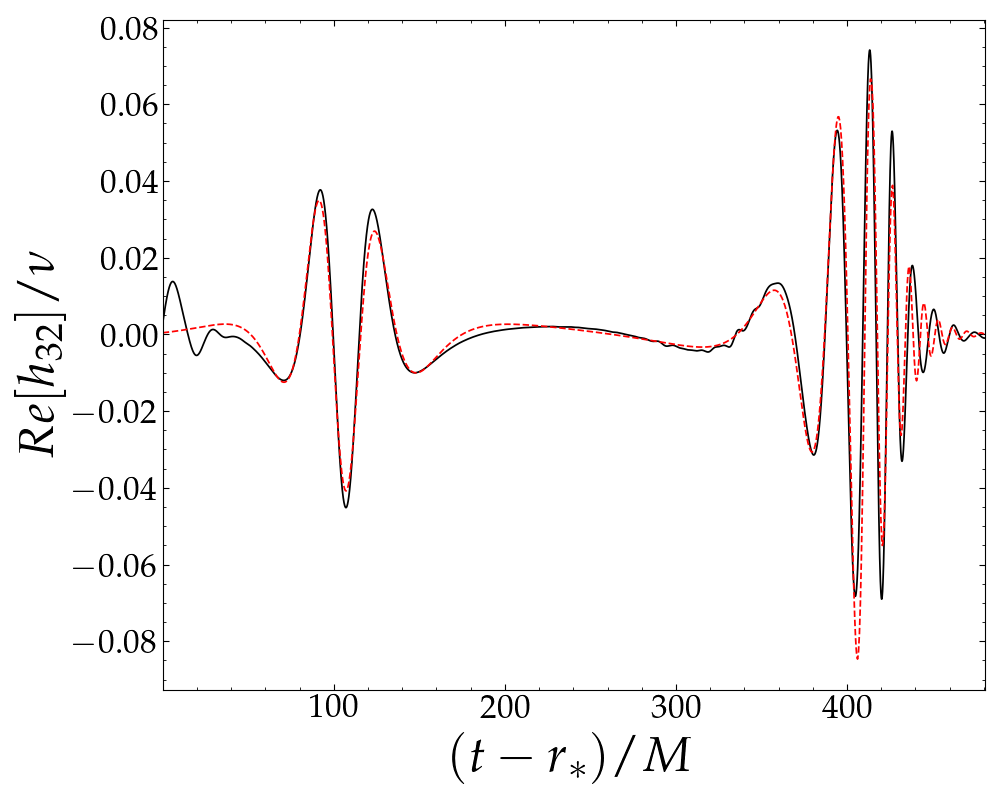}
       \includegraphics[width=.3\textwidth]{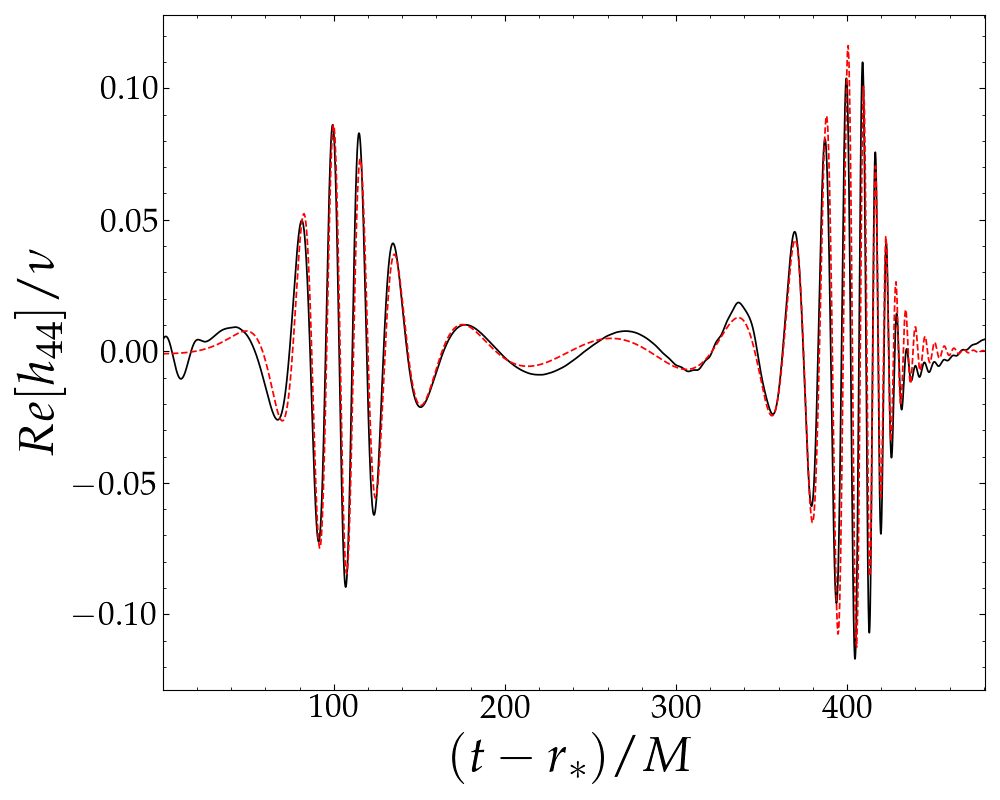}
    \includegraphics[width=.3\textwidth]{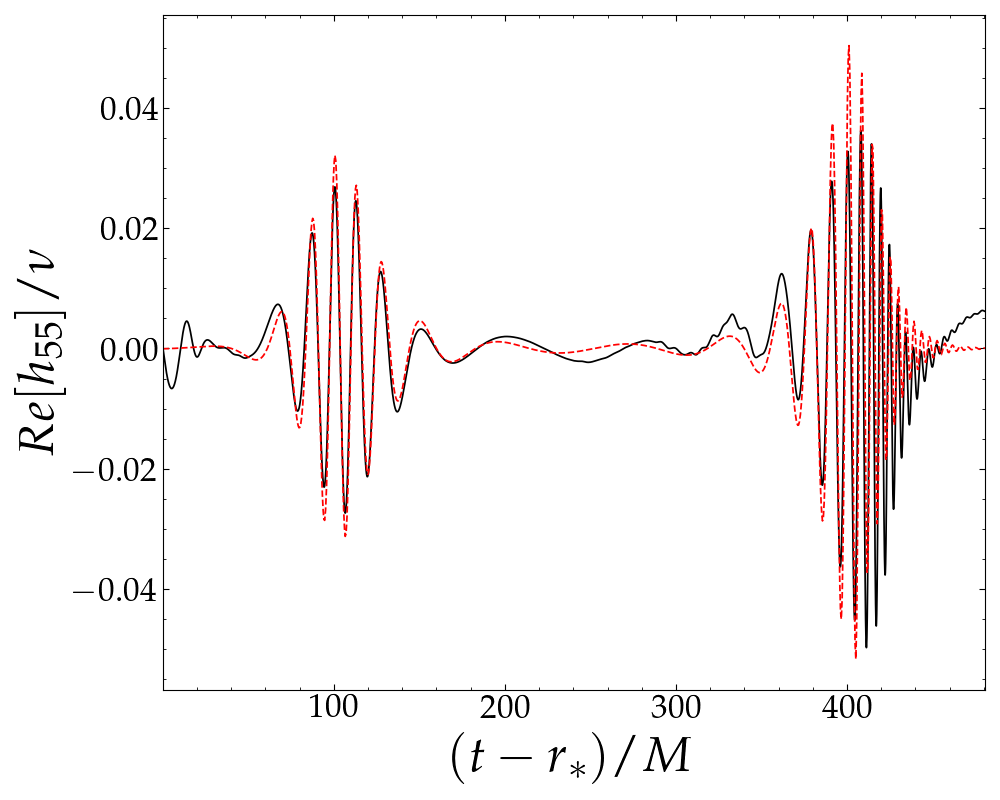}
    \caption{Unequal-mass ($q=2.15$), nonspinninga case, dataset \etksim{42q215}: EOB/NR phasing
    comparison including the available higher harmonics. Note that the $(2,2)$ mode is completed by
    the NR-informed, noncircular, merger and ringdown, while the higher harmonics still retain the standard
    quasi-circular contributions. Despite this, the agreement is acceptable. Also note some unphysical features
    in the ringdown for $(4,4)$ and $(5,5)$ due to the calculation of the strain from $\psi_4$.}
    \label{fig:HM} 
\end{figure*}

\begin{figure}[h] 
\centering
\includegraphics[width=.5\textwidth]{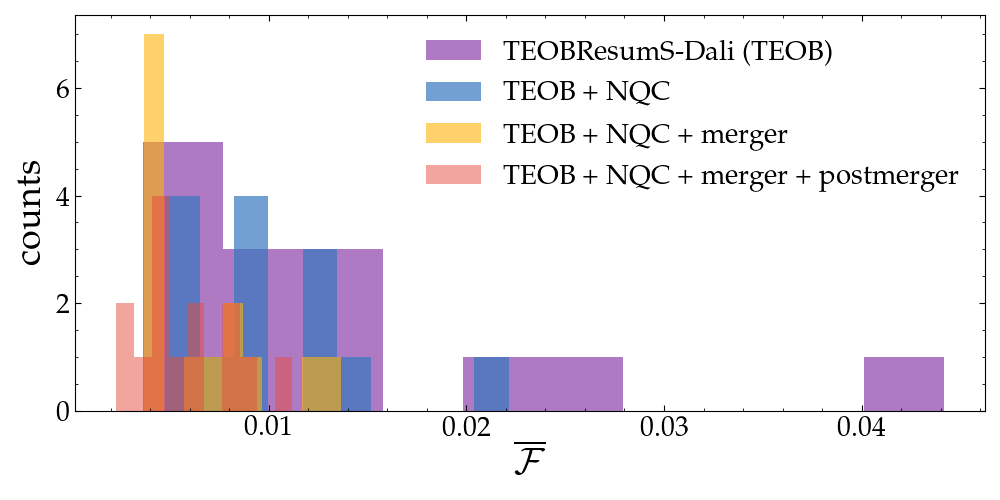}
\caption{EOB/NR unfaithfulness for the $\ell=m=2$ mode obtained in zero noise for all
configurations considered. Note how $\bar{\cal F}$ decreases due to the NR-improvement 
of the merger and ringdown part of the EOB waveform.}
\label{fig:mm}
\end{figure} 

\begin{figure}[h] 
\centering
\includegraphics[width=.5\textwidth]{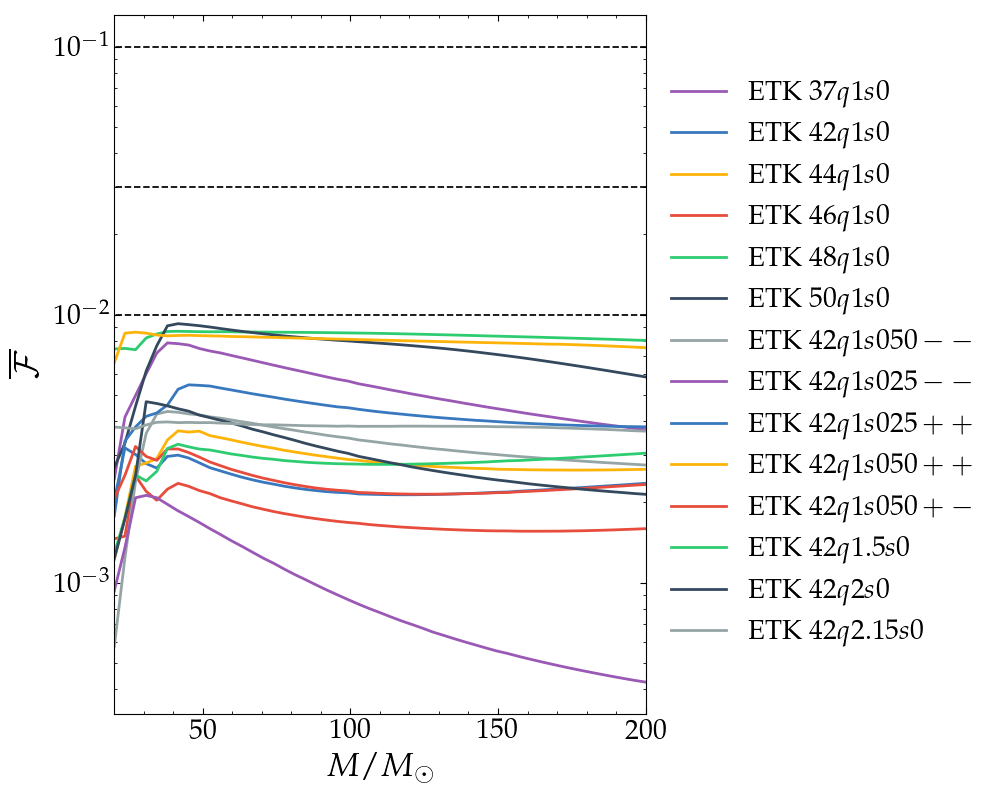}
\caption{EOB/NR unfaithfulness for the $\ell=m=2$ mode obtained with the zero-detuned, high-power 
noise spectral density of Advanced LIGO~\cite{aLIGODesign_PSD}.}
\label{fig:mm_psd}
\end{figure}

\begin{figure*}[t]
    \includegraphics[width=.38\textwidth]{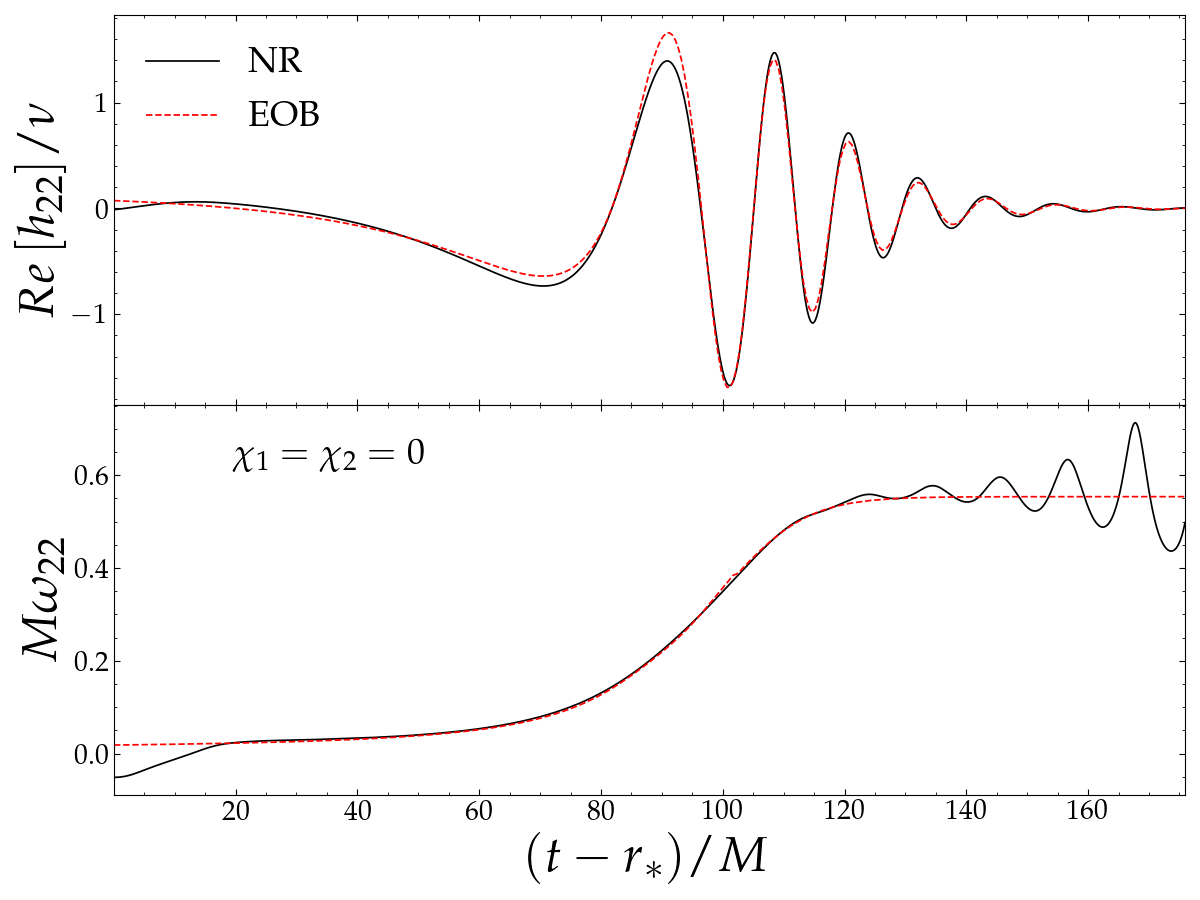}
    \hspace{5mm}  											
    \includegraphics[width=.38\textwidth]{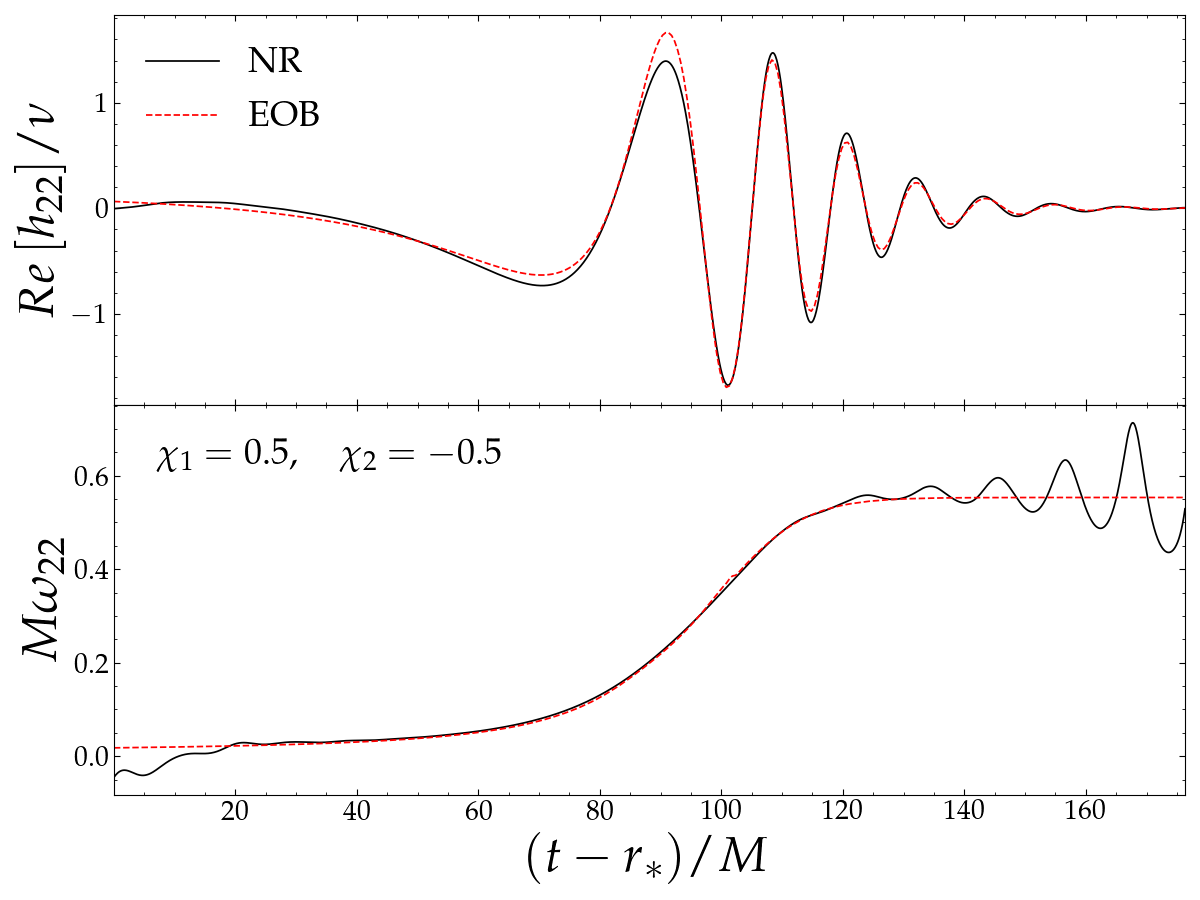}\\
    \vspace{5mm}
    \includegraphics[width=.38\textwidth]{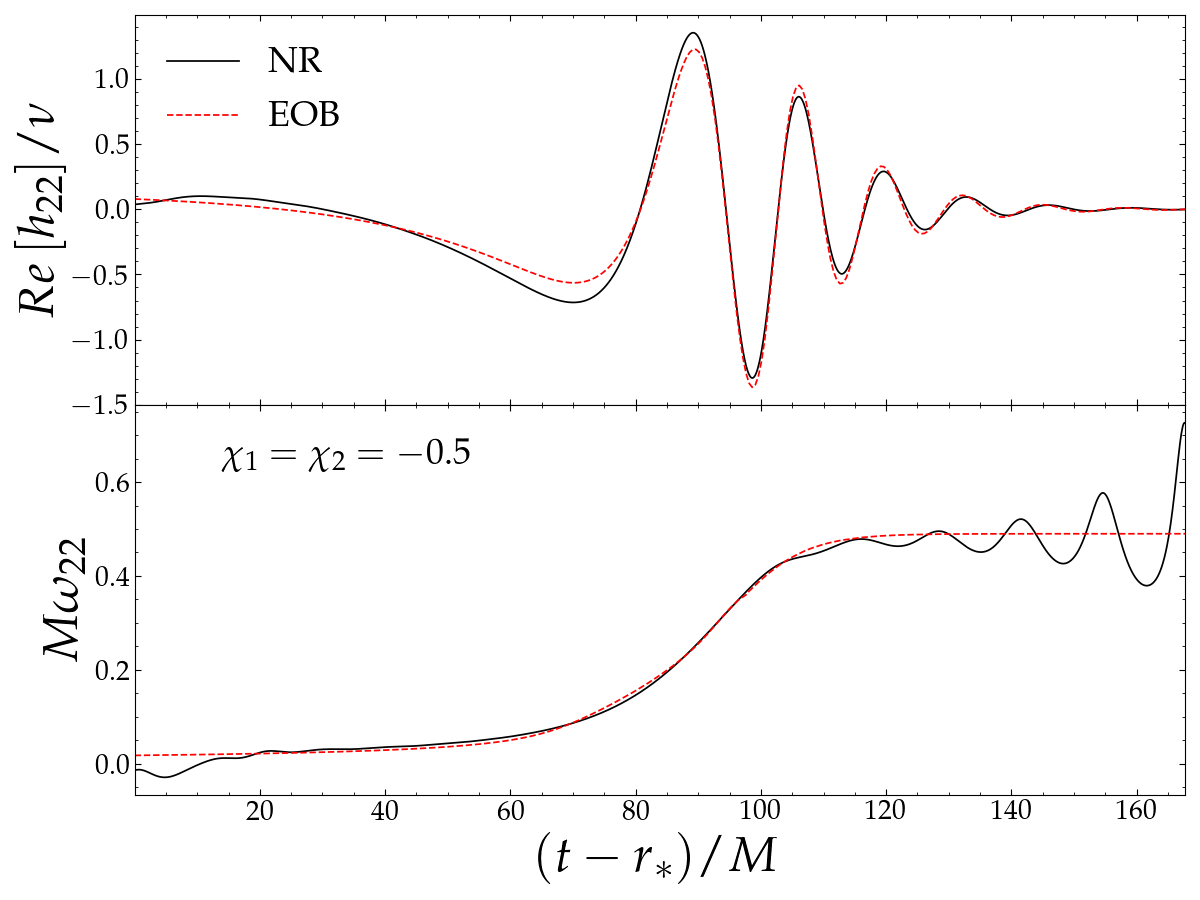}
    \hspace{5mm}
    \includegraphics[width=.38\textwidth]{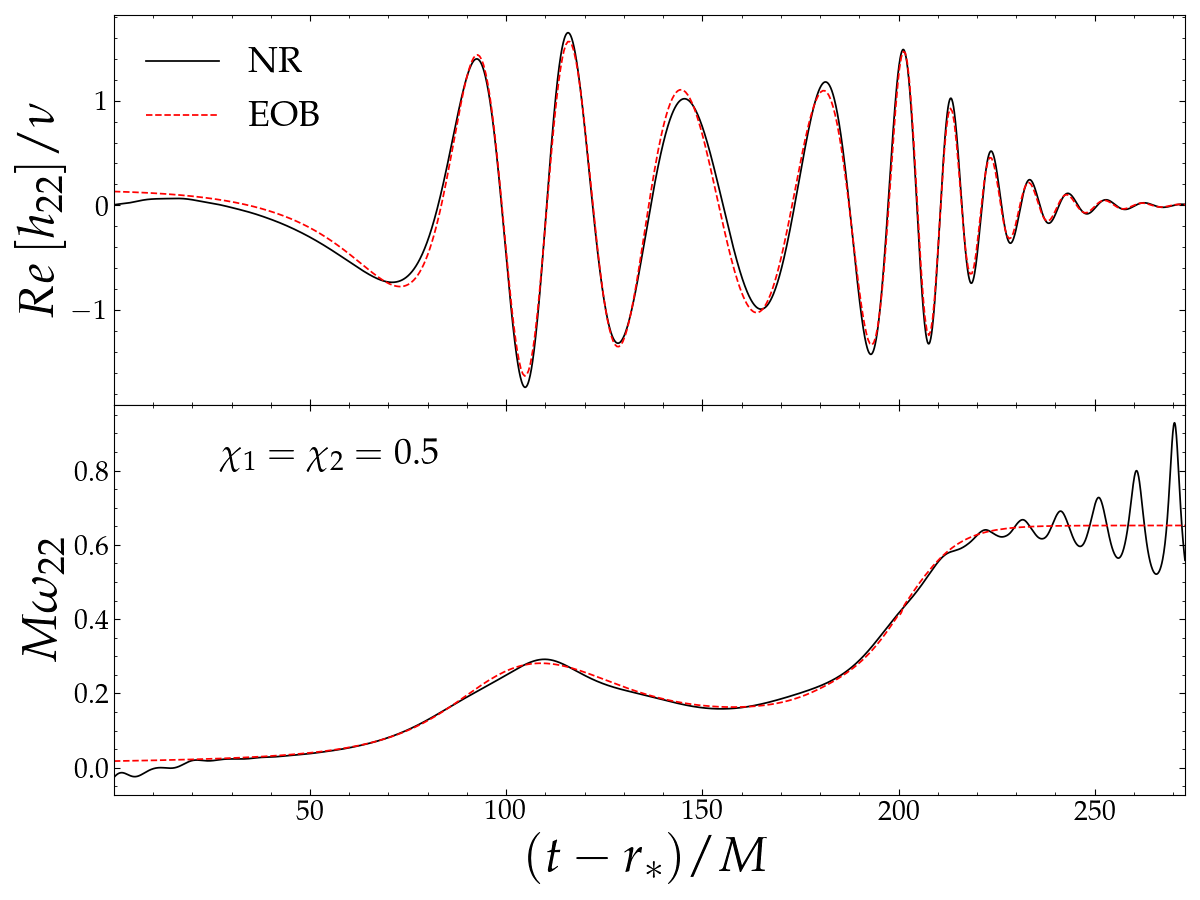}
      \caption{EOB/NR comparisons for spin-aligned binaries. Top row: \etksim{42q1s}* configurations with 
    $\chi_1=\chi_2=0$ (left) and $\chi_1=-\chi_2=0.5$ (right). The cancelation of the spin-orbit interaction as
    predicted by the EOB model, is present also in the NR data. Bottom panels: $\chi_1=\chi_2=-0.5$ (left) and
    $\chi_1=\chi_2=+0.5$ (right). When spins are negative, the plunge occurs faster and the signal
    is shorter than the $\chi_1=\chi_2=0$ case. When spins are positive, the repulsive character of 
    the spin-orbit coupling yields a first encounter followed then by the plunge and merger. 
    The phase differences are reported in Fig.~\ref{fig:Dphi}.}
    %
    \label{fig:spin05} 
\end{figure*}

\section{EOB/NR comparisons}
\label{sec:eobnr}
Let us finally focus on the bulk of our results about direct comparisons between EOB
and NR waveforms. As usual, we provide two different metrics: (i) phase and amplitude
differences in the time domain; (ii) EOB/NR unfaithfulness as defined from 
Eq.~\eqref{eq:F} above. In addition, we will also discuss comparisons between the
EOB and NR dynamics, as expressed using the gauge-invariant relation between
energy and angular momentum.

\subsection{Time-domain phasing and unfaithfulness}
The EOB/NR amplitude and phase differences are defined
\begin{align}
\label{Deltas}
    \Delta A^{\rm EOBNR}_{22} \equiv A^{\rm EOB}_{22} - A^{\rm NR}_{22} \ , \\
    \Delta \phi^{\rm EOBNR}\equiv \phi^{\rm EOB}_{22} - \phi^{\rm NR}_{22} \ ,
\end{align}
and are computed once the two waveforms are aligned after fixing an arbitrary
time and phase shifts $(\tau,\alpha)$. This is done by minimizing the unfaithfulness
(in zero noise) as defined from Eq.~\eqref{eq:F} above, computed using
the algorithm \texttt{optimized{\_}match} of 
the {\tt pyCBC} library~\cite{alex_nitz_2022_6912865}.

 \subsubsection{Nonspinning configurations} 
 \label{sec:nospin}
Let us discuss first nonspinning configurations, i.e. datasets  from~\etksim{31q1s0} to~\etksim{50q1s0}
(equal-mass) and \etksim{42q1.5s0} to \etksim{42q2.15s0} (unequal-mass). Focusing first on equal-mass case binaries,
the phenomenology changes from a direct capture, {\tt ETK31q1s0},  to a double encounter, \etksim{50q1s0}. 
The corresponding EOB/NR time-domain phasing comparisons are shown in Fig.~\ref{fig:nr_eob_42_50_all}. 
The top panels show the real part of the waveform and the instantaneous gravitational wave frequency, 
as obtained using the full NQC and ringdown improved model. The bottom panels exhibit the phase difference 
and the relative amplitude differences obtained with four different versions of the waveform:(i) the standard,
\TEOB{Dalì} one with the native ringdown informed by quasi-circular information;(ii) the model with NR-informed
NQC corrections;(iii) the model with NR-informed NQC corrections and NR-informed merger values 
of $(\hat{A}_{\rm max},\omega_{\rm max})$; (iv)the full improved model corresponding at the top panel, 
completed by the NR-informed complete postmerger waveform. The phase differences at merger corresponding
to case (iv) above are listed in Table~\ref{tab:eob_parameters}.
The bottom panels of Fig.~\ref{fig:nr_eob_42_50_all} indicate that the NR-information injected in the merger-ringdown
description may bring a reduction of the order of $\sim 0.1$~rad of the phase difference during the final phases of 
the coalescence. The differences during either the precursor (for~\etksim{31q1s0}) or the first 
encounter (for~\etksim{50q1s0}) are of the order of $0.1$~rad, with trends that suggest that some additional
analytic improvement, either in the dynamics or in the waveform~\cite{Albanesi:2022xge}, 
might  be needed to reach the NR accuracy level. Despite this, the corresponding values of 
the EOB/NR unfaithfulness $\bar{F}$ are more than acceptable, as we will see below.
Before presenting this calculation, let us focus on the few, nonspinning, dataset with $q\neq 1$.
The purpose of this choice was to reliably extract from NR also higher waveform modes 
and use them to test the corresponding EOB multipoles, tested so far only for eccentric
inspirals~\cite{Nagar:2021gss}. This is done in Fig.~\ref{fig:HM}, which shows the complete 
EOB/NR comparison for modes $\{(2,2),(2,1),(3,3),(3, 2), (4, 4), (5, 5)\}$ for the dataset
\etksim{42q2.15s0} with $q=2.15$. For this specific comparison we are using the standard
\TEOB{Dalì} model without NR-information {\it also} in the $(2,2)$ mode. Visually, the EOB/NR
agreement is acceptable and can be evidently improved further by injecting NR information.
Figure~\ref{fig:HM} also highlights inaccuracies in the recovering of the strain from $\psi_4$ from
higher modes, as it is evident from the trend of the post-merger waveform ffor modes $(4,4)$
and $(5,5)$.

The time-domain phasing analysis is complemented by the calculation of the unfaithfulness $\bar{\cal F}$.
This is done either assuming zero noise, $S_n(f)=1$, or using the zero-detuned, high power 
spectral density (PSD) design sensitivity of Advanced Ligo~\cite{aLIGODesign_PSD}. As it is standard
for quasi-circular binaries, this yields $\bar{\cal F}$ as function of the total mass.
Assuming $S_n(f)=1$, we explore how the increase of NR information used to correctly 
shape the merger-ringdown part of the waveform reflects on $\bar{F}$. The result of this analysis 
is found in Fig.~\ref{fig:mm}. The values corresponding to the 
complete model are also listed in the last column of Table~\ref{tab:ETK_sims}.
Note that the mismatches are highest for the smallest angular momenta simulation
(corresponding to lowest scattering angle \etksim{37q1s0} and anti-aligned spins \etksim{42q1s050+-},
a dataset to be discussed below).  In all other cases, the mismatches are below $1\%$, always
obtained after the NR-information procedure. The calculation using the Advanced LIGO PSD
is exhibited in Fig.~\ref{fig:mm} for total mass $20 M_\odot \leq M \leq 200 M_\odot$. 
It is remarkable to note that, despite the absence of any additional tuning on the actual dynamics of the 
binary (that was informed using only quasi-circular simulations), one has 
$\bar{\cal F}^{\max} \sim 1\%$ all over the considered configurations.

 \subsubsection{Spin}
 \label{sec:spin}
We also considered a few datasets with spin aligned (or anti-aligned) with
the orbital angular momentum. Before discussing their properties and
putting them in relation with the EOB model, let us recall a pure EOB prediction 
presented in Ref.~\cite{Nagar:2020xsk}. The effect of the BH spins (aligned or
anti-aligned with the orbital angular momentum) for dynamical capture BBHs 
was analyzed in Sec.~IIIA of Ref.~\cite{Nagar:2020xsk}. The spin-orbit interaction 
implemented within the EOB model allowed for a very precise prediction for a 
$q=1$ dynamical encounter of the changes in the waveform phenomenology 
with respect to the corresponding nonspinnig case. 
\begin{figure}[t]
   \includegraphics[width=.4\textwidth]{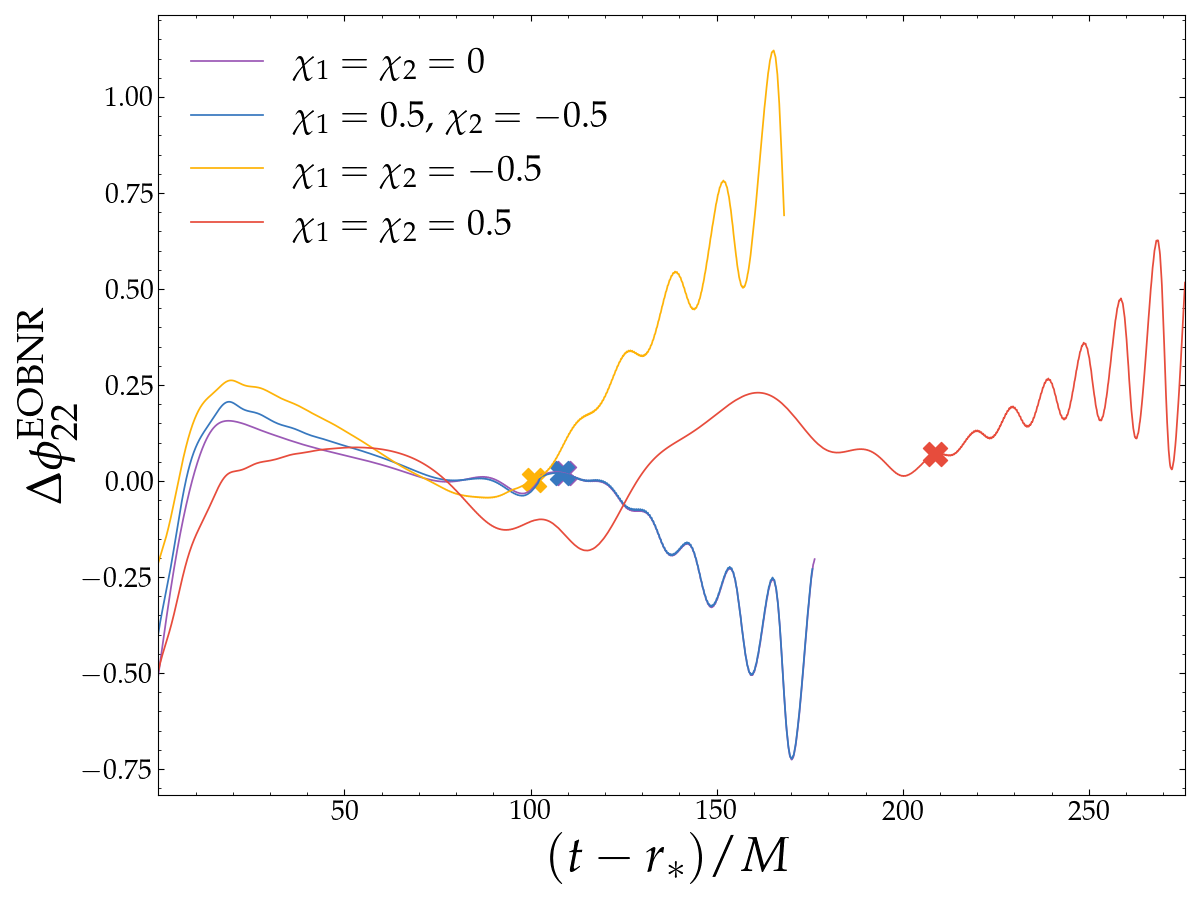} 
\caption{\label{fig:Dphi}EOB/NR phase differences for the configurations of Fig.~\ref{fig:spin05}.
The markers indicates the location of the waveform amplitude peak. Note the rather small
differences (only during the precursor phase) between the $\chi_1=\chi_2=0$ and 
$\chi_1=-\chi_2=0.5$ datasets.}
\end{figure}      
More precisely: (i) if spins are both aligned with angular momentum, 
the {\it repulsive} character of the interaction is such that a configuration that merge
in absence of spins can just scatter; (ii) if spins are both anti-aligned with the angular
momentum, the {\it attractive} character of the spin-orbit interaction entails the plunge
to occur on a shorter time scale; (iii) if spins are one aligned and one anti-aligned with
the orbital angular momentum the spin-orbit interaction cancels out and the corresponding
waveform is substantially equivalent to the nonspinning one.
This phenomenology is summarized in Fig.~6 of Ref.~\cite{Nagar:2020xsk}.
Here, we start from the nonspinning configuration \etksim{42q1s0} and add spins, with
dimensionless magnitude $|\chi|=0.25$ and $|\chi|=0.5$ and different orientations. 
In this second case, we consider the three possible configurations, 
$\chi_1=\chi_2=+0.5$, $\chi_1=\chi_2=-0.5$ ad $\chi_1=-\chi_2=+0.5$
so to test the analytical prediction of Ref.~\cite{Nagar:2020xsk} discussed above. 
Also here, as it was the case before, it is intended that the EOB waveform for
each configuration is completed by the NR-informed NQC corrections and ringdown.

The first row of Fig.~\ref{fig:spin05} displays the nonspinning simulation (left panel) and 
the one with misaligned spins (right panel). Both NR datasets are also compared with the 
corresponding EOB waveform (completed by the NR-informed description of merger and ringdown).
We see that the analytical prediction of Ref.~\cite{Nagar:2020xsk} is fully confirmed,
with just very small differences between the waveforms due to spin-spin effects. 
The bottom row of the picture shows the case with both spins negative (left panel) and
positive (right panel). In this second case, we see how the repulsive character of the 
spin-orbit interaction yields a first encounter (highlighted by the presence of a local
maximum in the gravitational wave frequency) then followed by the merger.
The actual EOB/NR phase differences are quantified in Fig.~\ref{fig:Dphi}, which depicts
together $\Delta\phi^{\rm EOBNR}_{22}$ for the four datasets. 

\subsection{Dynamics}
\label{sec:dyn}
We also compared the EOB and NR dynamics expressed using the gauge-invariant relation
between energy and angular momentum~\cite{Damour:2011fu,Nagar:2015xqa}. On the NR side, 
this quantity can be extracted as the parametric curves $(j(t), E_b(t))$ 
given by Eq.~\eqref{eq:Eb_and_j} for $t > t_{{\rm junk}}$.
On the EOB side, this is just obtained from the evolution of the Hamiltonain dynamics. 
Note in this respect that it is {\it not} computed from the waveform multipoles as in the NR case.
We gather the plots for all simulations in Appendix~\ref{app:NR_EOB_plots}, and list 
in Table~\ref{tab:eob_parameters} as meaningful values only the differences at merger. 
It is interesting to note that EOB/NR differences decrease monotonically with all physical 
parameters $J_i$, $q$, and $\chi$, i.e. as the signals become longer and thus closer to 
the quasi-circular simulations used to inform the model. 

\section{Conclusions}
\label{sec:conclusions}

We have presented new NR simulations of dynamical capture black hole binaries.
Our set of simulations includes some of the configurations of 
Gold and Br\"ugmann~\cite{Gold:2012tk} (see also~\cite{East:2012xq}),
improving the systematic exploration of aligned spins and mass ratio effects.
The runs were mostly done using the \ETK{} NR code. A few configurations were also
simulated using the \GRA{} code for mutual cross checking. The NR strain waveforms
(computed here for the first time) are first compared with the state-of-the-art 
eccentric model \TEOB{} and additionally used to inform it to obtain improved 
accuracy for merger and ringdown. Our results on the NR side can be summarized 
as follows:
\begin{itemize}
\item[(i)]We have systematically analyzed configurations with spins. 
We found that the analytical EOB predictions due to spin-orbit interaction of 
Ref.~\cite{Nagar:2020xsk} in spin-aligned encounters (see Fig.~ \ref{fig:spin05}) are fully confirmed 
by NR simulations. More precisely: if the spins are positive, the system has more GW cycles 
before merger than in the nonspinning case (repulsive character of spin-orbit interaction); 
if the spins are both negative, the plunge occurs faster with less cycles (attractive character of 
spin-orbit interaction); when the spins are misaligned and equal (one positive and one negative) 
the dynamics and waveforms are fully compatible with the nonspinning
case.

\item[(ii)]We presented the first direct comparisons between \ETK{} and \GRA{}, corresponding to
three different nonspinning equal-mass dynamical captures, previously studied in 
Ref.~\cite{Gold:2012tk}. The comparison between the codes shows good quantitative agreement 
(Sec.~\ref{sec:code_cf}), and the independent codes display self-convergence (Sec. ~\ref{sec:self-conv}). 
\item[(iii)]We have systematically computed the strain waveform (that was absent in Ref.~\cite{Gold:2012tk})
for both \ETK{} and \GRA{}, increasing the amount of currently known information (see in particular the
supplementary material of Ref.~\cite{Gamba:2021gap}, where 3 more configurations obtained with \GRA{}
were presented). The computation of the strain from $\psi_4$ is one of the main technical challenging
aspect of these simulations, and gets worse for higher modes. These difficulties point to the 
need of directly extracting the strain at infinity from NR simulations using well known 
techniques based on Regge-Wheeler-Zerilli perturbation theory~\cite{Abrahams:1995gn, Nagar:2005ea,Pazos:2006kz, Buonanno:2009qa} 
or even using Cauchy Characteristic Extraction~\cite{Bishop:2009zz, Reisswig:2009us, Reisswig:2009rx}.
See also \cite{CalderonBustillo:2022dph} for an alternative way of performing parameter inference that uses $\psi_4$ directly.
\item[(iv)]We computed, for the first time, the relation between energy and angular momentum, extending
work previously done only for quasi-circular binaries~\cite{Damour:2011fu,Nagar:2015xqa,Ossokine:2017dge}.
\end{itemize}
On the EOB side, our results can be summarized as follows
\begin{itemize}
\item[(i)]To start with, we compare the 14 $\ell=m=2$ NR waveforms obtained with 
the waveforms obtained using the state-of-the-art EOB model  \TEOB{-Dalì}. The computation
of the EOB/NR mismatch in white noise are $~1.5\%$ for all datasets except for those configurations
with the lowest angular momentum, that reach $\sim 4\%$. This is by itself a remarkable result considering
that  \TEOB{-Dalì} was only informed by quasi-circular NR simulations~\cite{Nagar:2021gss}.
\item[(ii)]We report of good agreement also for higher modes. Notably, for double encounter configurations,
this is true also for the first burst of radiation. This indicates that the accuracy of the multipolar Newtonian
prefactor in the waveform, introduced in Ref.~\cite{Chiaramello:2020ehz}, and only tested 
for bound configurations~\cite{Nagar:2021gss}, is maintained also for hyperbolic encounters.
\item[(iii)]We then used NR simulations to improve the EOB model, notably the $\ell=m=2$ 
merger and ringdown part. We found that the accuracy of the EOB ringdown is mostly dominated 
by the values of amplitude and
frequency at merger and of the mass and spin of the final black hole. When these values are incorporated
in the model, as well as NR-informed NQC corrections, the EOB/NR mismatches are at most $1.1\%$ for all 
the simulations considered.  
\end{itemize}
Our analysis lays out the procedure of informing \TEOB{-Dalì} incorporating NR data, showcasing that the model is sufficiently accurate for future searches of dynamical encounters signals. Future work will focus on a systematic numerical investigation of dynamical encounters and present global fits of the merger and ringdown parameters to inform \TEOB{-Dalì}. We thus expect to obtain a highly faithful extension of \TEOB{-Dalì} for parameter estimation of dynamical capture and highly eccentric events, which will allow for further reassessment of the events observed by LVK so far.


\section*{Acknowledgments}

We thank  Juan Garc\'ia-Bellido and Santiago Jaraba for insightful discussions on NR simulations of hyperbolic encounters and related matters, and Mark Gieles for valuable comments on rates of eccentric mergers. 
We thank all the participants of the ``Workshop on Gravitational Wave Modelling'' held in Barcelona (2022) and of the ``EOB@Work 2023'' workshop in Jena.
The work of TA is supported in part by the ERC Advanced Grant GravBHs-692951 and 
by Grant CEX2019-000918-M funded by Ministerio de Ciencia e Innovaci\'on (MCIN)/Agencia 
Estatal de Investigaci\'on (AEI)/10.13039/501100011033.
RG acknowledges support from the Deutsche Forschungsgemeinschaft
(DFG) under Grant No. 406116891 within the Research
Training Group RTG 2522/1.
SB knowledges support from the EU H2020 under ERC Starting
Grant, no.~BinGraSp-714626, from the EU Horizon under ERC Consolidator Grant,
no. InspiReM-101043372.
JCB is supported by a fellowship from ``la Caixa'' Foundation (ID
100010434) and from the European Union's Horizon 2020 research and innovation programme under the Marie Sklodowska-Curie grant agreement No 847648. The fellowship code is LCF/BQ/PI20/11760016. JCB is also supported by the research grant PID2020-118635GB-I00 from the Spain-Ministerio de Ciencia e Innovaci\'{o}n.
NSG is supported by the Spanish Ministerio de Universidades, through a María Zambrano grant (Grant No. ZA21-031) with reference UP2021-044, within the European Union-Next Generation EU.
JAF is supported by the Spanish Agencia Estatal de Investigaci\'on (Grants PGC2018-095984-B-I00 and PID2021-125485NB-C21) funded by MCIN/AEI/10.13039/501100011033 and ERDF A way of making Europe, by the Generalitat Valenciana (Grant PROMETEO/2019/071), and by the European Union’s Horizon 2020 research and innovation (RISE) programme H2020-MSCA-RISE-2017 Grant No. FunFiCO-777740.
DR work was supported by NASA under award No.~80NSSC21K1720.
We are grateful to P.~Micca for, never trivial, inspiring suggestions.

\noindent \TEOB{} is publicly available at {\small {\url{https://bitbucket.org/eob_ihes/teobresums/}}} \\

\noindent The {\ETK} simulations were performed at Lluis Vives cluster for scientific 
calculations at University of Valencia~\footnote{
\url{https://www.uv.es/uvweb/servei-informatica/ca/serveis/investigacio/calcul-cientific-universitat-1286033621557.html}} 
and MareNostrum4 at the Barcelona Supercomputing Center (Grants No. AECT-2022-3-0001 and AECT-2022-2-0006).
The {\GRA} simulations were performed on
the national HPE Apollo Hawk 
at the High Performance Computing Center Stuttgart (HLRS), 
on the ARA cluster at Friedrich Schiller University Jena 
and on the supercomputer SuperMUC-NG at the Leibniz-Rechenzentrum 
(LRZ, \url{www.lrz.de}) Munich. The ARA cluster is funded 
in part by DFG grants INST 275/334-1 FUGG and INST 275/363-1 
FUGG, and ERC Starting Grant, grant agreement no.~BinGraSp-714626. 
The authors acknowledge HLRS for funding this project by 
providing access to the supercomputer HPE Apollo Hawk under 
the grant number INTRHYGUE/44215. The authors acknowledge 
also the Gauss Centre for Supercomputing e.V. 
(\url{www.gauss-centre.eu}) for funding this project by 
providing computing time to the GCS Supercomputer SuperMUC-NG 
at LRZ (allocation pn68wi).

\appendix

\section{Puncture tracks and $\psi_4$}
\label{app:trajectories}
\begin{figure*}[t] 
\centering
\includegraphics[scale=0.4]{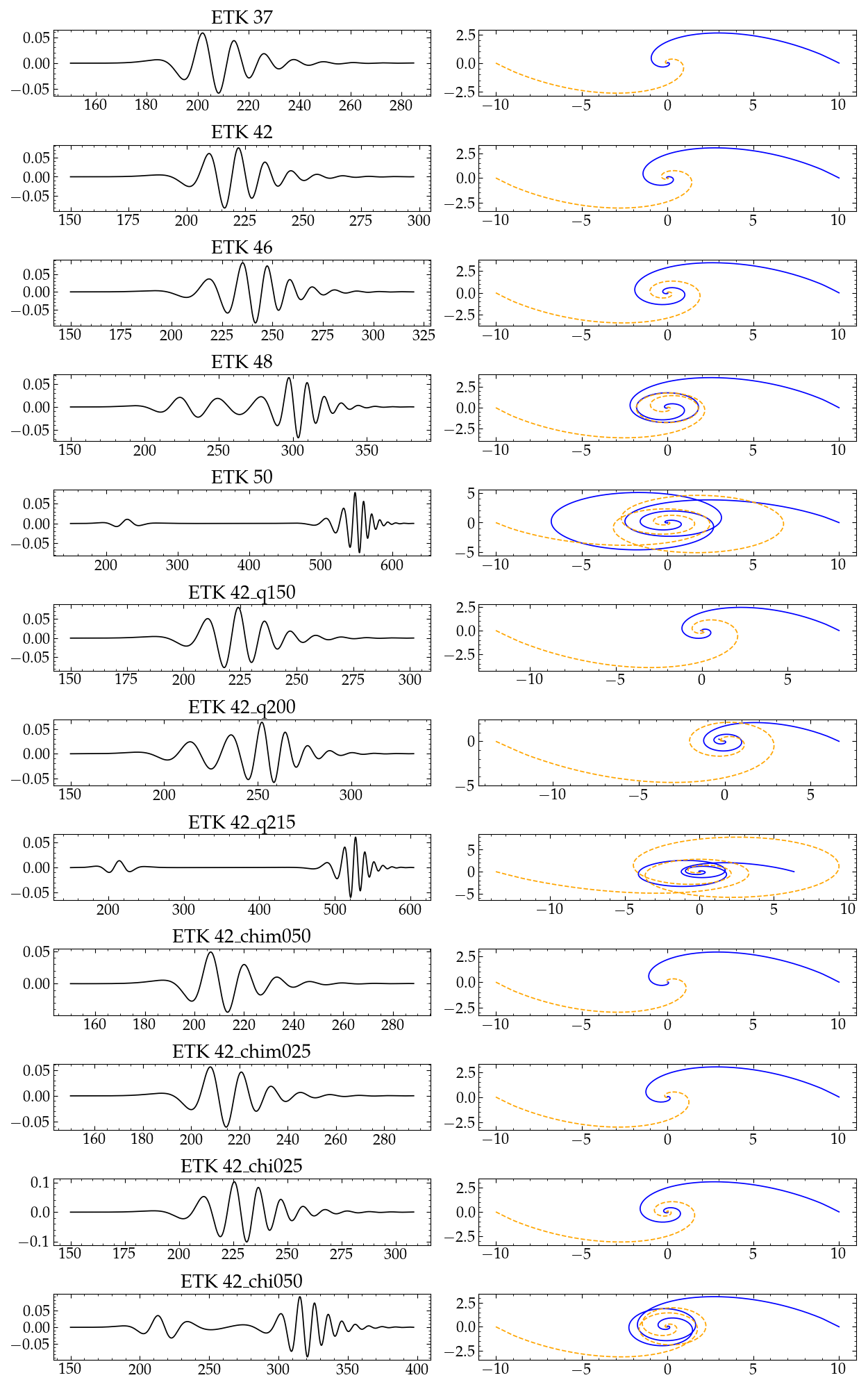}
\caption{Puncture tracks and $\psi_4^{(2,2)}$ for all \ETK{} simulations}
\label{fig:trajectories}
\end{figure*}

In this Appendix we gather the unprocessed data for 
$\psi_4$ extracted at $R = 100 M$ and the corresponding 
puncture trajectories for all simulations. This information is
collected in Fig.~\ref{fig:trajectories}.
%

\section{NR technical information}

\subsection{Code comparison for $\psi_4$}
\label{app:cc_psi4}

\begin{figure*}[]
  \includegraphics[width=.8\textwidth]{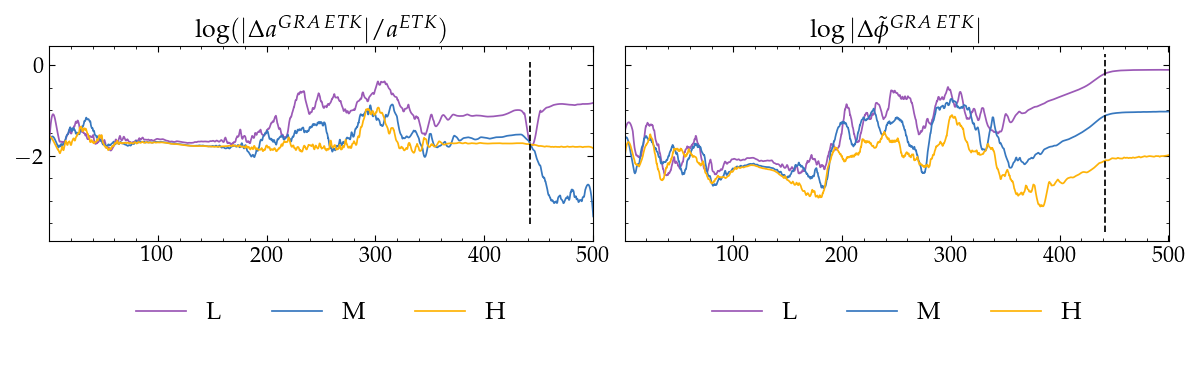}   
  \caption{\label{fig:code_comparison_psi4} Comparison between {\ETK} and {\GRA} 
  simulations \etksim{50} and \athsim{50} at fixed resolution. 
  We display the absolute value of the difference in normalized 
  amplitude and phase of the non-extrapolated $\psi_4$ for every available resolution. 
  }
\end{figure*}

This appendix shows the comparison between the leading modes of $\psi_4$ obtained from 
{\ETK} and {\GRA} simulations {\it without} using extrapolation. Note that this is the data obtained 
directly from our numerics, so this is a direct comparison between both codes. 
We write the leading mode of $\psi_4$ in terms of its phase and shift by  
\begin{equation}
\label{eq:psi_4_amp_phase}
    \psi_4(t) = a(t) e^{i \tilde \phi(t)}
\end{equation}
We show our results for the differences in Fig. \ref{fig:code_comparison_psi4}, and gather the results for the differences at merger in Table \ref{tab:code_comparison_psi4}. 

\begin{table}[h] 
\caption{\label{tab:code_comparison_psi4} [Amplitude and phase differences at merger for the
leading mode of $\psi_4$ in simulations  \etksim{42q1s0,48q1s0,50q1s0}, \athsim{42q1s0,48q1s0,50q1s0} carried out at low, medium, and high resolutions, as listed in Table \ref{tab:self_conv_info}.]}
\begin{center}
\begin{ruledtabular}
\begin{tabular}{ c c | c c } 

ID & res & $\Delta a_{{\rm mrg}}^{GRA \,ETK} /a^{ETK}$ & $\Delta \tilde \phi_{{\rm mrg}}^{GRA \,ETK}$ \\ 
 \hline 
 \hline 
 &L & $ 9.610  \cdot 10^{-3} $ & $ 2.179  \cdot 10^{-2} $ \\ 
{\tt 42q1s0} &M & $ 1.612  \cdot 10^{-2} $ & $ 1.833  \cdot 10^{-2} $ \\ 
 &H & $ 1.728  \cdot 10^{-2} $ & $ 1.535  \cdot 10^{-2} $ \\ 
\hline 
 &L & $ 1.826  \cdot 10^{-2} $ & $ 1.592  \cdot 10^{-1} $ \\ 
{\tt 48q1s0} &M & $ 1.963  \cdot 10^{-2} $ & $ 2.253  \cdot 10^{-2} $ \\ 
 &H & $ 1.834  \cdot 10^{-2} $ & $ 5.373  \cdot 10^{-3} $ \\ 
\hline 
 &L & $ 1.817  \cdot 10^{-2} $ & $ 6.526  \cdot 10^{-1} $ \\ 
{\tt 50q1s0} &M & $ 1.791  \cdot 10^{-2} $ & $ 7.651  \cdot 10^{-2} $ \\ 
 & H & $ 1.761  \cdot 10^{-2} $ & $ 7.139  \cdot 10^{-3} $
\end{tabular}
\end{ruledtabular}
\end{center}
\end{table}

\section{Extra EOB/NR plots}
\label{app:NR_EOB_plots}

In this appendix we gather plots for the EOB/NR comparisons for the leading modes of the strain for 
all of our {\ETK} simulations. The EOB model includes all of the NR information discussed in the main 
text, i.e. ring-down, and NQC corrections. We show our results for the real part of $h$, the modes 
frequencies, and the binding energy versus dimensionless angular momentum.

\begin{figure*}[] 
\centering
\includegraphics[scale=0.22]{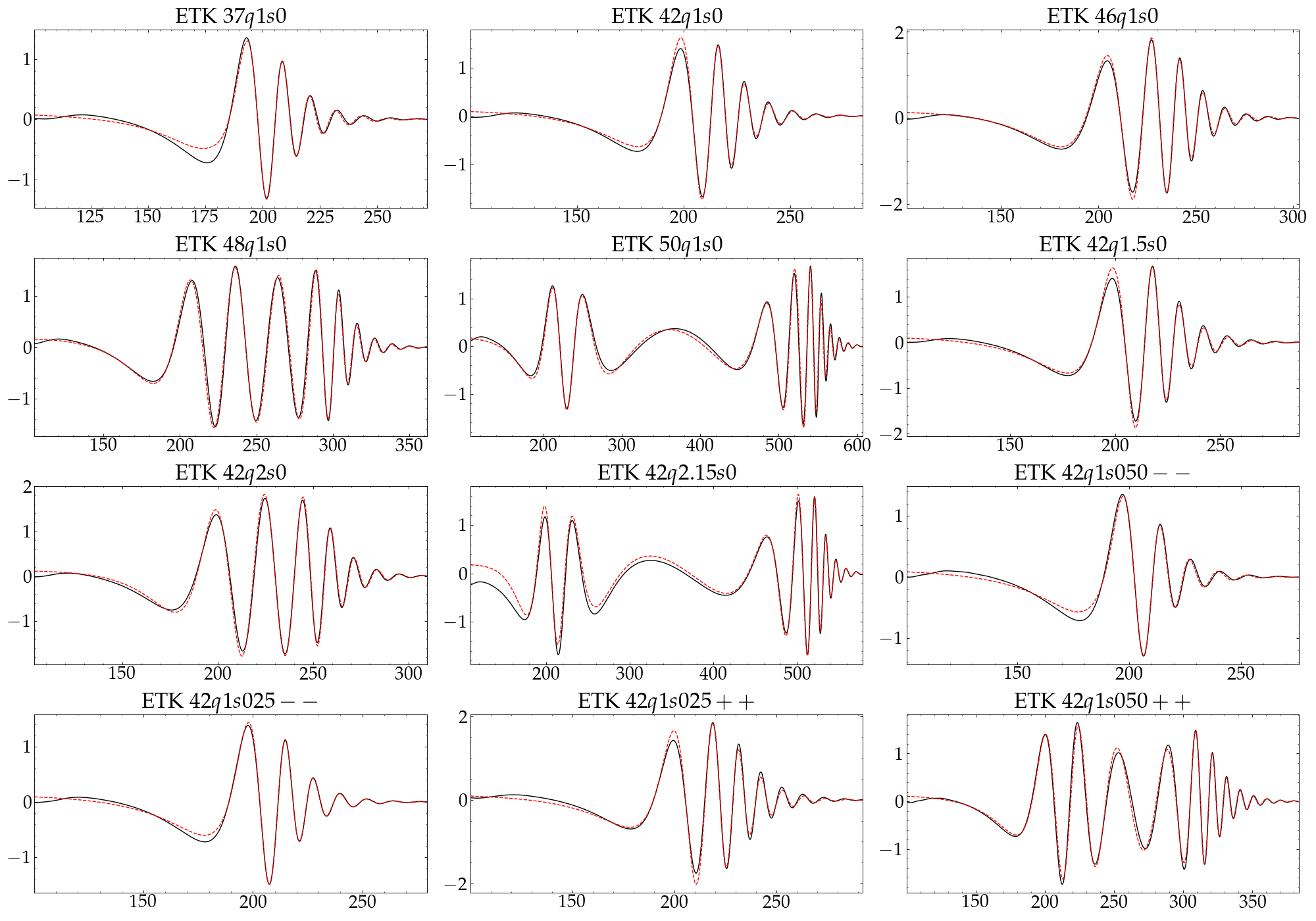}
\includegraphics[scale=0.4]{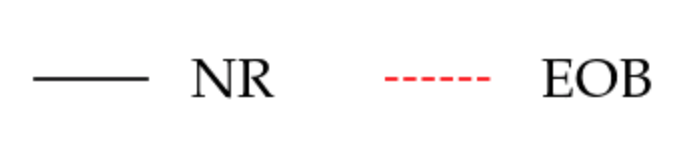}
\caption{EOB/NR comparison plots for the real parts of the waveforms for the leading $(2,2)$ modes, 
normalized by $\nu$. In the EOB case, we plot the fully NR-informed model, including ringdown and NQC parameters.
}
\label{fig:nreob_real_tile}
\end{figure*} 
\begin{figure*}[] 
\centering
\includegraphics[scale=0.25]{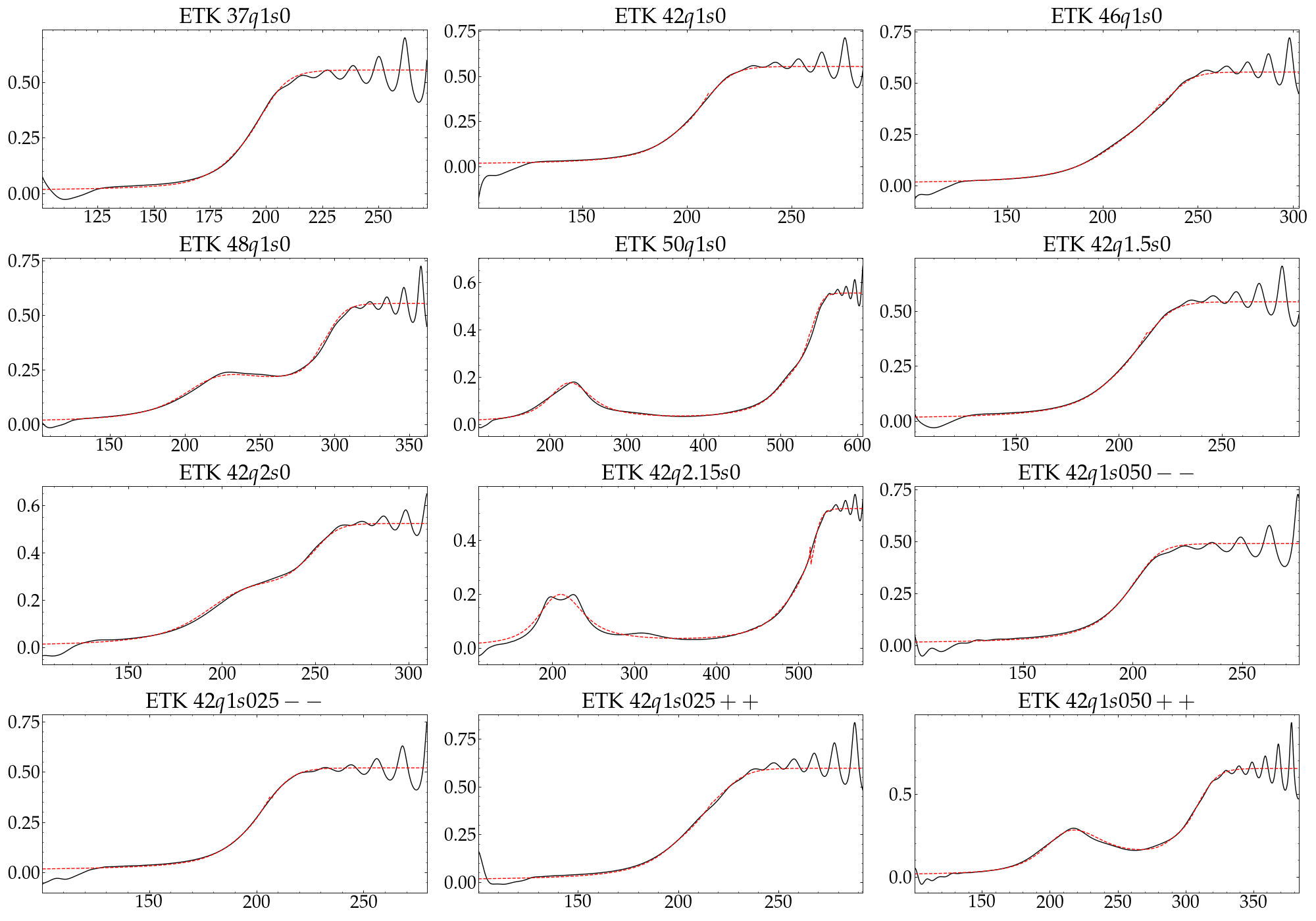}
\includegraphics[scale=0.35]{fig/nr_eob_label.png}
\caption{EOB/NR comparison plots for the frequency of the leading $(2,2)$ modes.
In the EOB case, we plot the fully NR-informed model, including ringdown and NQC parameters.}
\label{fig:nreob_freq_tile}
\end{figure*} 
\begin{figure*}[] 
\centering
\includegraphics[scale=0.25]{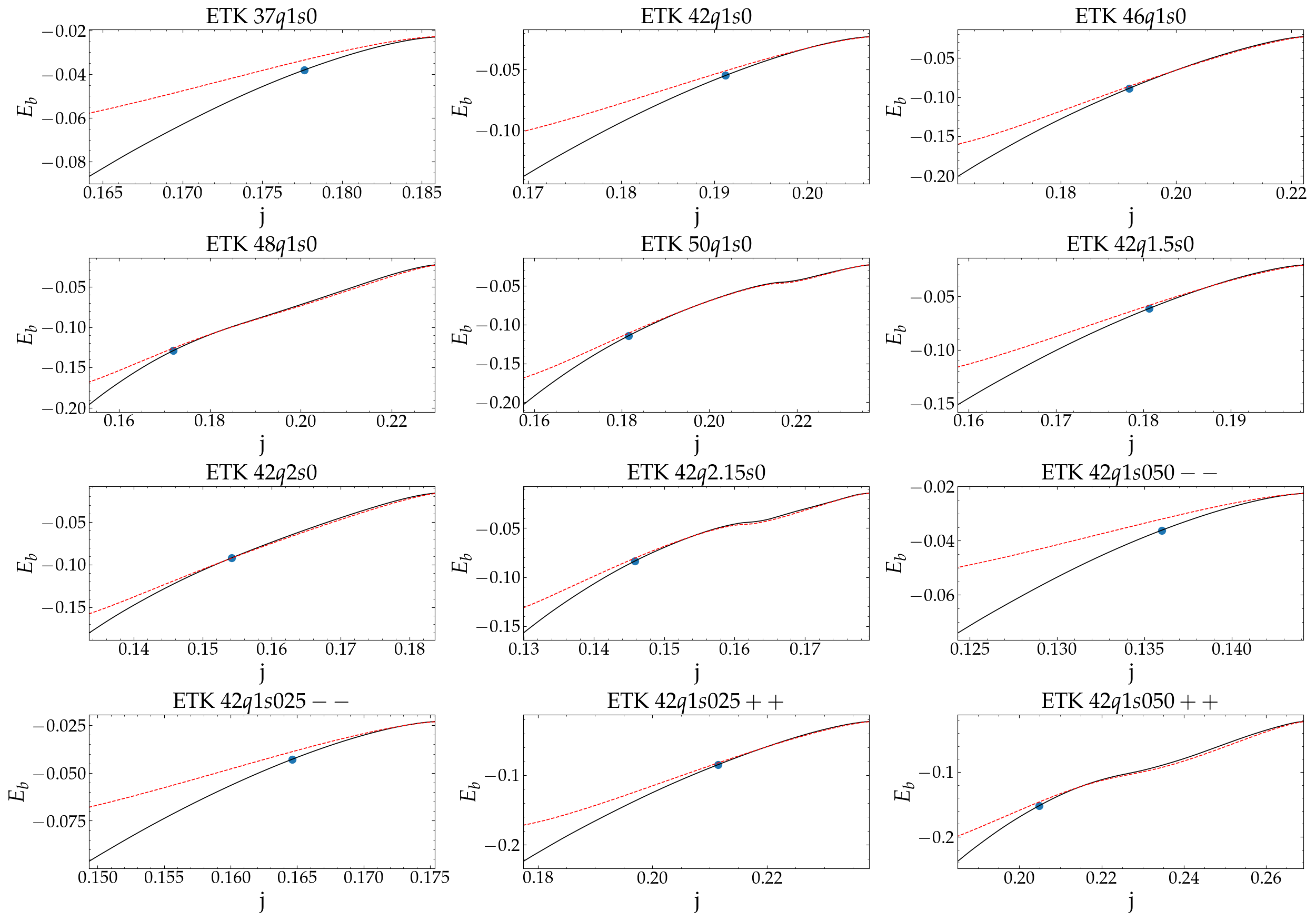}
\includegraphics[scale=0.35]{fig/nr_eob_label.png}
\caption{Binding energy versus dimensionless angular momentum curves for all simulations. 
The blue marker denotes the NR merger value for each dataset. 
}
\label{fig:EJ_curves}
\end{figure*} 

\bibliography{refs.bib,local.bib}

\end{document}